\documentclass{aa}  

\usepackage{graphicx}
\usepackage{txfonts}

\begin{document}

   \title{A method for identifying metal-poor stars with \newline Gaia BP/RP spectra \thanks{Table A.1 and a table with all the coefficients of the method are only available in electronic form at the CDS via anonymous ftp to cdsarc.u-strasbg.fr (130.79.128.5) or via http://cdsweb.u-strasbg.fr/cgi-bin/qcat?J/A+A/}}

   \author{T. Xylakis-Dornbusch
          \inst{1,\thanks{Member of the International Max Planck Research School for Astronomy \& Cosmic Physics at the University of Heidelberg (IMPRS-HD).}},
          N. Christlieb\inst{1}, K.Lind\inst{2} \and T.Nordlander\inst{3,4}
          }

   \institute{Zentrum f\"ur Astronomie der Universit\"at Heidelberg, Landessternwarte, K\"onigstuhl 12, 69117 Heidelberg, Germany\\
              \email{txylaki@lsw.uni-heidelberg.de}
         \and
             Department of Astronomy, Stockholm University, AlbaNova University Center, SE-106 91 Stockholm, Sweden\\
        \and
            Research School of Astronomy and Astrophysics, Australian National University, Canberra, ACT 2611, Australia\\
        \and
            ARC Centre of Excellence for All Sky Astrophysics in 3 Dimensions (ASTRO 3D), Australia\\
             }

   \date{Received }

 
  \abstract
   {The study of the oldest and most metal-poor stars in our Galaxy promotes our understanding of the Galactic chemical evolution and the beginning of Galaxy and star formation. However, they are notoriously difficult to find, with only five stars at $\mathrm{[Fe/H]<-5.0}$ having been detected to date. Thus, the spectrophotometric data of 219 million sources which became available in the third Gaia Data Release comprise a very promising dataset for the identification of metal-poor stars.}
   {We want to use the low-resolution Gaia Blue Photometer / Red Photometer (BP/RP) spectra to identify metal-poor stars. Our primary aspiration is to help populate the poorly constrained tail of the metallicity distribution function of the stellar halo of the Galaxy.}
   {We developed a metal-poor candidate selection method based on flux ratios from the BP/RP Gaia spectra, using simulated synthetic spectra.   
   }
   {We found a relation between the relative iron abundance and the flux ratio of the Ca H \& K region to that of the $\mathrm{H\beta}$ line. This relation is temperature and surface gravity dependent, and it holds for stars with $\mathrm{4800\,K \leq T_{eff}\leq6300\,K}$. We applied it to noisy simulated synthetic spectra and inferred $\mathrm{[Fe/H]}$ with an uncertainty of $\sigma_{\mathrm{[Fe/H]}}\lessapprox0.65$ dex for $\mathrm{-3\leq[Fe/H]}\leq 0.5$ and G=15-17mag, which is sufficient to identify stars at $\mathrm{[Fe/H]<-2.0 }$ reliably. We predict that by selecting stars with inferred $\mathrm{[Fe/H]}\leq-2.5$ dex, we can retrieve 80\% of the stars with $\mathrm{[Fe/H]}\leq-3$ and have a success rate of about 50\%, that is one in two stars we select would have $\mathrm{[Fe/H]}\leq-3$. We do not take into account the effect of reddening, so our method should only be applied to stars which are located in regions of low extinction.}
   {}

    \titlerunning{Prospects for identifying metal-poor stars with Gaia DR3 BP/RP spectra}
    \authorrunning{T.Xylakis et al.}
 
   \maketitle
%

\section{Introduction}
The Gaia survey \citep{2016A&A...595A...1G} poses an unprecedented opportunity to understand the structure, history, and evolution of our Galaxy better. Aside from the astrometry and photometry for over a billion stars, Gaia has also delivered spectrophotometric data in its third Data Release (Gaia DR3), which was made public on June 13, 2022. This could serve as a tool to map out the Galaxy in terms of the relative iron abundance $\mathrm{[Fe/H]}$  -- which is a proxy for the total amount of metals in a star -- and consequently find new metal-poor stars. Stars deficient in elements heavier than helium, called metal-poor stars, are usually very old, with ages comparable to that of the Universe \citep{Frebel_2013}. The basic idea is that stellar atmospheres preserve, to a large extent, the makeup of their birth cloud, hence a metal deficient star should have formed along with the first few generations of stars after the Big Bang. That is why these objects are very interesting, and they can facilitate our understanding of the past, even though they can be found close by. In particular, they can be used as tracers of the evolution of the elements and of the Galaxy, help us understand how the distribution of the first population of stars (Pop III) looked, and assist in constraining Big Bang nucleosynthesis further  \citep{https://doi.org/10.1002/asna.201011362,doi:10.1146/annurev.astro.42.053102.134057,2008ARA&A..46..241S}. \newline
 The Gaia Collaboration itself, through the Data and Analysis Consortium (DPAC), has already, and will also in the future, deliver astrophysical parameters using the Gaia data. With the astrometric and photometric data of Gaia DR2, \cite{2018A&A...616A...8A} delivered temperatures, extinctions and reddening. DPAC also used the Gaia DR3 Blue Photometer / Red Photometer (BP/RP) spectra to estimate metallicities of stellar objects \citep{2022arXiv220605864C,2022arXiv220605992F,2022arXiv220606138A}. 
 \cite{2012MNRAS.426.2463L} predicted that Aeneas -- which is a Bayesian method employed by DPAC for the stellar parameter estimation -- would estimate the metallicity with an accuracy of 0.1-0.2 dex, given that the extinction is $\mathrm{A_{0}<2\,mag}$ and that the magnitude is $G=15 $mag. Now with Gaia DR3 available, \cite{2022arXiv220606138A} estimated $\mathrm{[M/H]}$ values for several million sources, and provide information on how to use them.\newline
 The search for metal-poor stars is as pertinent as ever now. The ongoing photometric SkyMapper Southern Sky (SMSS) survey is actively searching for extremely metal-poor stars ($\mathrm{[Fe/H]<-3\,dex}$ according to the classification of \cite{doi:10.1146/annurev.astro.42.053102.134057}). With its DR1.1, the SkyMapper team found that 40\% of their candidates had $\mathrm{[Fe/H]<-2.75\,dex}$ and only $\approx7\%$ had $\mathrm{[Fe/H]>-2\,dex}$, that is an efficacy of $\approx93\%$ in finding stars with $\mathrm{[Fe/H]<-2\,dex}$ \citep{2019}. SkyMapper also led to the discovery of the star with the lowest detected Fe abundance \citep{2019MNRAS.488L.109N}. The \textit{Pristine} survey, which is being carried out in the northern hemisphere, was tailored to find metal-poor stars with a narrow-band filter centered on the metallicity-sensitive Ca II H \& K lines \citep{2017}. \textit{Pristine} has a 56\% reported efficacy of finding stars with $\mathrm{[Fe/H]<-2.5\,dex}$ and 23\% for stars with $\mathrm{[Fe/H]<-3\,dex}$ \citep{2019pristine}. The objective of this work is to develop a recipe that can efficiently identify metal-poor stars, and especially stars within and below the extremely metal-poor regime ($\mathrm{[Fe/H]<-3\,dex}$). In Section \ref{section1} we present the reasoning of our approach for this endeavor, as well as the tools we used. We also describe the parameter space where our method is applicable, along with a detailed description of the development of our technique. Further, in Section \ref{sec_results} we present our results from applying our method on noise-free and noisy spectra. Therein, we also investigate the dependence of our procedure on the different astrophysical parameters, and we additionally study the effect of extinction. Lastly, we explore the expected efficiency of our technique.

\section{Methods}\label{section1}
\begin{figure}
    \centering
    \includegraphics[width=0.5\textwidth]{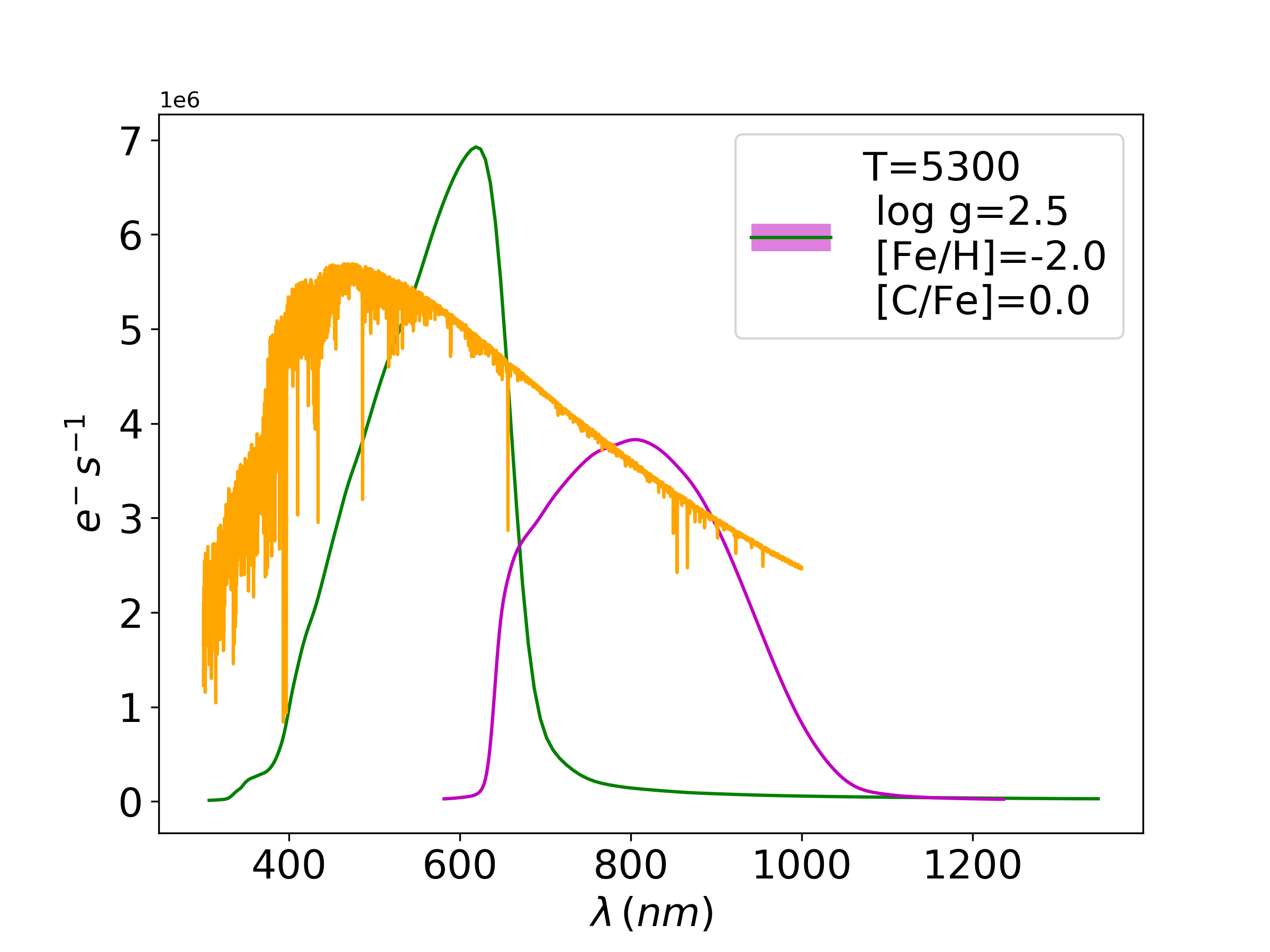}
    \caption{Synthetic spectrum before (orange) and after (green and magenta) simulation. The spectrum has $\mathrm{T_{eff}=5300\,K,\,\log g=2.5\,dex,\, [Fe/H]=-2 \,dex}$ and $\mathrm{[C/Fe]=0.0\,dex}$.}
    \label{fig:before_after}
\end{figure}

For this work, we used the Ulysses Simulator \citep{Ulysses} and synthetic spectra \citep{2019MNRAS.488L.109N} (see Sections \ref{subsubsection2} and \ref{subsubsection1}, respectively) in order to simulate the spectrophotometric data of BP and RP, respectively, on board Gaia. The very low resolution of the expected BP/RP spectra (see Figure \ref{fig:before_after}) inclined us to use integrated fluxes of different parts of the spectra, such that a relation varying with metallicity could be found. Those spectral areas had to be sensitive to the change in metallicity and also to at least one other quantity that can be known a priori or that can be directly derived from the spectra themselves.

 \subsection{Data and simulations}\label{subsection1}
\begin{figure}
    \centering
    \includegraphics[width=0.4\textwidth]{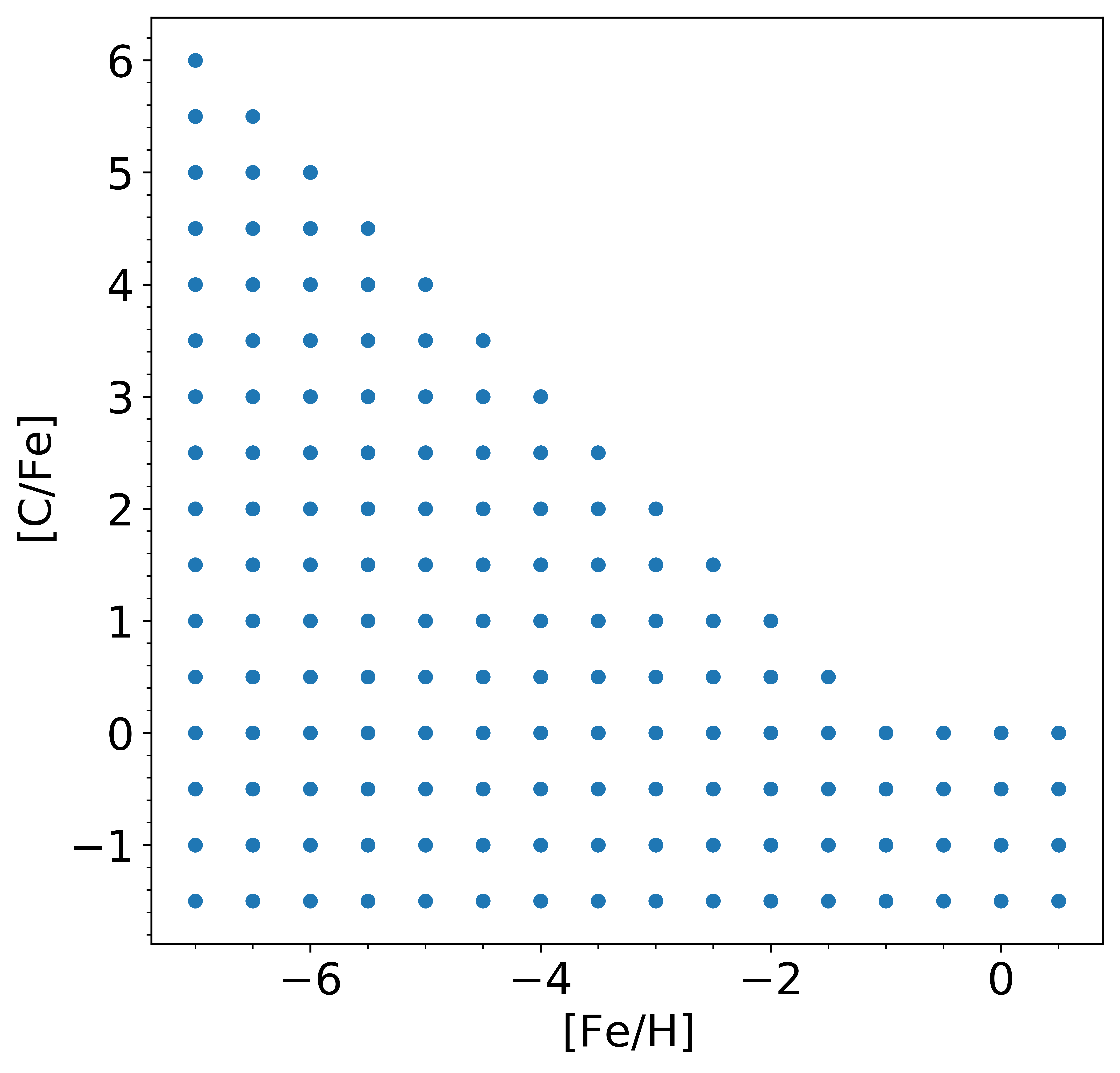}
    \label{fig:feh_cfe_combination}
    \includegraphics[width=0.4\textwidth]{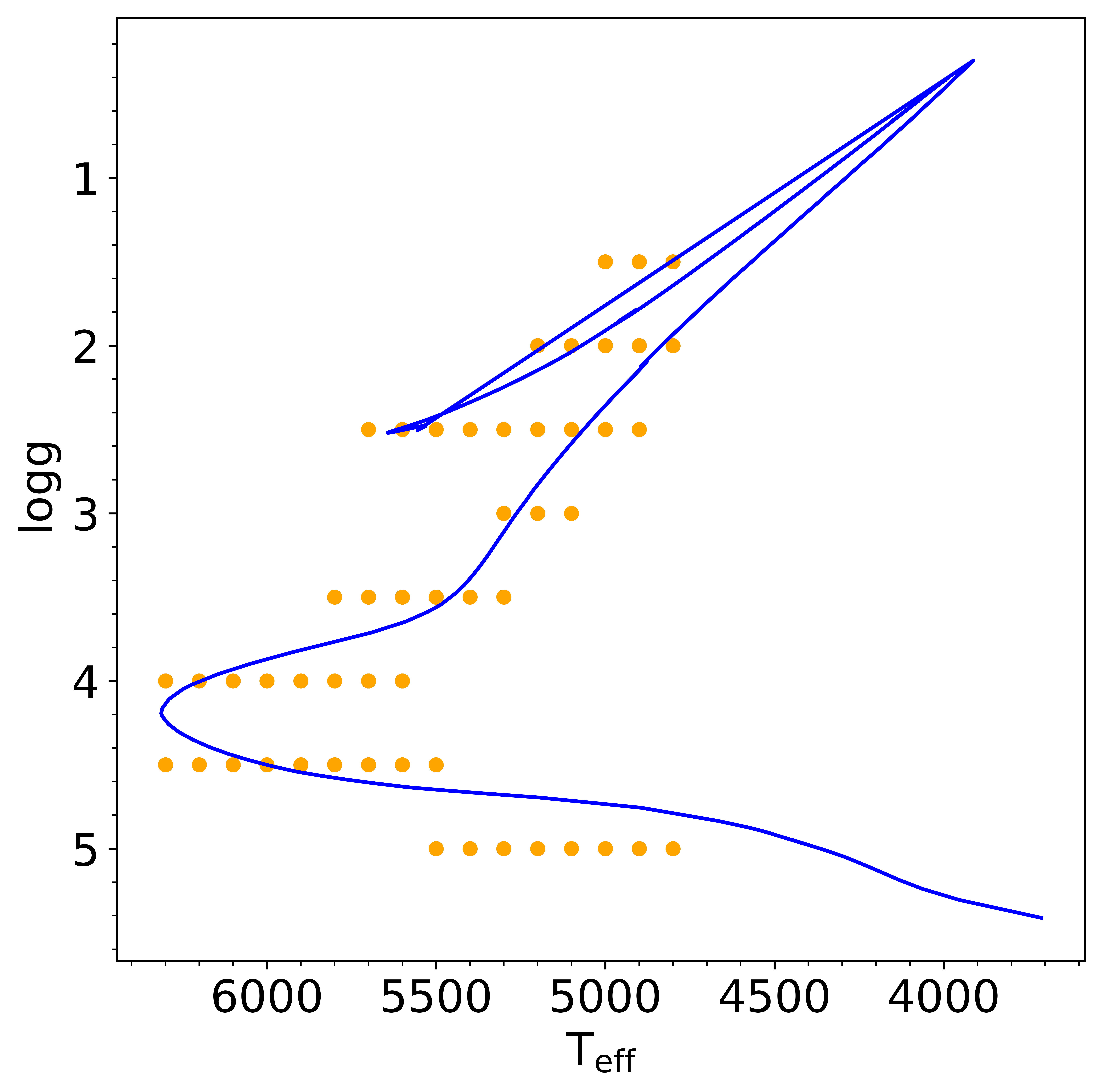}
    \caption{Parameter space we used to develop our method. Top plot: Each point in the top plot represents a specific combination of relative carbon and iron abundances. Our entire dataset is comprised of all the meaningful temperature and surface gravity combinations (see bottom plot) for each and every one of these relative abundance combinations. Bottom plot: A PARSEC isochrone of 12 Gyr and $\mathrm{[Fe/H]=-3.0\,dex}$, which we used to determine the astrophysical parameters of our preliminary dataset. Overplotted are the temperature-surface gravity pairs of the final parameter space we used to develop our method.}
   \label{fig:isochrone_params}
\end{figure}

 Choosing the temperature and surface gravity space for our dataset was the first critical step for our method. The desired relation has to hold in that parameter space, which would then allow us to use those parameters as priors when applying the method described in this paper. Additionally, the dataset has to cover a wide metallicity range, so that both metal-poor and metal-rich stars can be covered, and we can ultimately distinguish between them. Since low metallicities are involved in the process,, the last important factor that we have to pay attention to is the carbon enhancement that has been observed in many metal-poor stars \citep{doi:10.1146/annurev.astro.42.053102.134057,2015review,2021arentsen}. The initial dataset parameters are shown in Table \ref{table:ssp_param_space} and Figure \ref{fig:feh_cfe_combination}.

 \begin{table}[!ht]
        \caption{Initial parameter space of the synthetic spectra.}
        \label{table:ssp_param_space}
    \centering
    \begin{tabular} {c c c}
    \hline\hline
    
    Parameter & Range & Step\\
    \hline\\
  $\mathrm{T_{eff}}$ (K) &  $3500\,\leq \mathrm{T_{eff}}\leq 7800$ & 100 \\
    $\mathrm{\log g}$ (dex) & $0.5\leq \mathrm{\log g}\leq 5.0$ & 0.5 \\
    $\mathrm{[Fe/H]}$  (dex) & $-7.0\leq \mathrm{[Fe/H]}\leq 0.5$& 0.5 \\
    $\mathrm{[C/Fe]}$  (dex) & $-1.5\leq \mathrm{[C/Fe]}\leq 6.0$ & 0.5  \\
    \hline
    
    \end{tabular}
    \tablefoot{The metallicity and carbon enhancement combinations are restricted to physically meaningful combinations (see Figure \ref{fig:feh_cfe_combination}).}
\end{table}

\subsubsection{Synthetic spectra}\label{subsubsection1}
For this work, we used synthetic spectra from a one dimensional (1D) grid where local thermodynamic equilibrium (LTE) is assumed ~\citep{2019MNRAS.488L.109N}, that was computed with the Turbospectrum code ~\citep{1998A&A...330.1109A, 2012ascl.soft05004P} and MARCS model atmospheres ~\citep{2008A&A...486..951G}. \cite{2019MNRAS.488L.109N} adopted the Solar chemical composition from \cite{2009ARA&A..47..481A}; for $\mathrm{[Fe/H]<-1.0}$, $\mathrm{[\alpha/Fe]}=+0.4$ was adopted; for models with $\mathrm{-1.0<[Fe/H]<0.0}$, a relationship of $\mathrm{[\alpha/Fe]=+0.4\cdot[Fe/H]}$ was used. The models take continuum scattering into account and use $v_{mic}=1\,\mathrm{kms^{-1}}$ for models with $\mathrm{\log g}\geq4.0$. Models with $\mathrm{\log g}\leq3.5$ take spherical symmetry into account and assume $v_{mic}=2\,\mathrm{kms^{-1}}$. The line data are a combination of atomic data from VALD3 \citep{2015PhyS...90e4005R} and molecular data from a variety of sources; here the primary molecular line list is the one for CH from \cite{2014A&A...571A..47M}. This grid also contains spectra with varying carbon abundances, which we used for this work. 

\subsubsection{Ulysses}\label{subsubsection2}
Ulysses \citep{Ulysses} is a simulator which takes spectra as input and delivers the equivalent BP/RP spectrum as it would be observed from Gaia. The final DR3 BP/RP spectra will be a combination of many different epoch observations, spanning up to the entire time of the missions' operation. That is why the input spectrum is being convolved with an averaged line spread function (LSF). For each of the two spectrophotometers (BP and RP), an averaged LSF exists, which is the product of all the LSFs of the different CCDs and telescopes. The parameters of Ulysses that can be tuned, among others, are the number of transits -- which are the number of times an object was observed -- the amount of interstellar extinction $A_0$, the apparent magnitude of the source, as well as the spectrum oversampling. The spectrum oversampling is essentially the number of spectra that will be combined into the final BP/RP spectrum. Each spectrum is sampled over 60 pixels, but for different observations, different parts of the spectrum are being sampled, so that there will be more than 60 pixels of data in the end. \cite{2021A&A...652A..86C} present a model for the internal calibration of the BP/RP spectra, in particular how to produce a mean spectrum from all the epoch spectra of the same source. The extinction curve models that Ulysses implements are those of \cite{1989ApJ...345..245C} and \cite{1999PASP..111...63F}. We selected the \cite{1999PASP..111...63F} models for our simulations, after determining both models produce very similar results. \newline The core products of Ulysses are noise-free BP/RP spectra, noise-free Gaia photometry, and extinctions. Other outputs are also possible, for instance the end-of-mission noisy spectra, which are being generated with the addition of noise to the noise-free spectra. The noise model employed by Ulysses is the one from \cite{2010A&A...523A..48J}. An overview of the configuration we used to simulate our data is shown in Table \ref{table:ulysses}.

\subsubsection{Parameter space}\label{subsubsection3}
A $\chi^2$ test of the simulated spectra showed that they carry enough information in the region of the Ca II H \& K lines to distinguish between metal-rich stars and stars with $\mathrm{[Fe/H]<-2.0}$. 
However, this was observed only for stars with $\mathrm{T_{eff}\geq4800\,K}$ and also depending on their surface gravity. A $\mathrm{12\,Gyr}$ and $\mathrm{Z=0.001}$ isochrone \citep{2012MNRAS.427..127B} was used to choose realistic $\mathrm{T_{eff}}$ - $\mathrm{\log g}$ pairs for the preliminary dataset (see Figure \ref{fig:isochrone_params}). 

\begin{table}
        \caption{Configuration of the Ulysses simulator which we used in order to construct our method.}
        \label{table:ulysses} 
        \centering
    \begin{tabular} {c c}
    \hline\hline
    Parameter & Value \\\hline\\
    \multicolumn{1}{l}{G magnitude (mag)} & 15  \\
    \multicolumn{1}{l}{spectrum oversampling} & 4 \\
    \multicolumn{1}{l}{number of transits} & 75 \\
    \multicolumn{1}{l}{extinction $\mathrm{A_0}$} & 0.0 \\ \hline
    \end{tabular}
    \tablefoot{The method is described in Section \ref{subsubsection2}. The same configuration was used for the test and applications described in Section \ref{sec_results}, apart from the magnitude and extinction which are explicitly mentioned when they have different values than in the table above.}
\end{table}

\subsection{Flux ratios}\label{subsection2}
\begin{figure}
   \centering
   \includegraphics[width=0.5\textwidth]{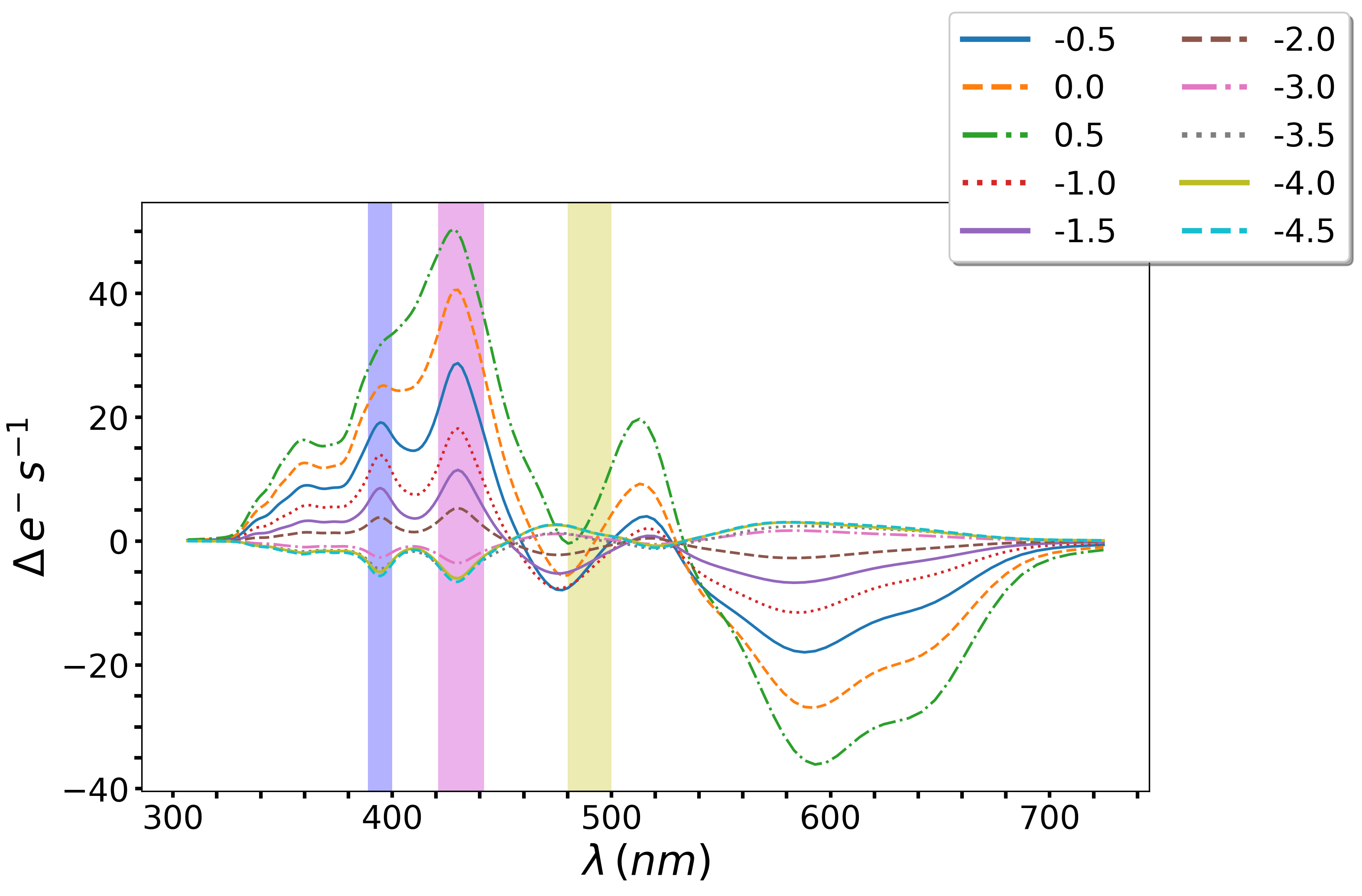}
    \includegraphics[width=0.5\textwidth]{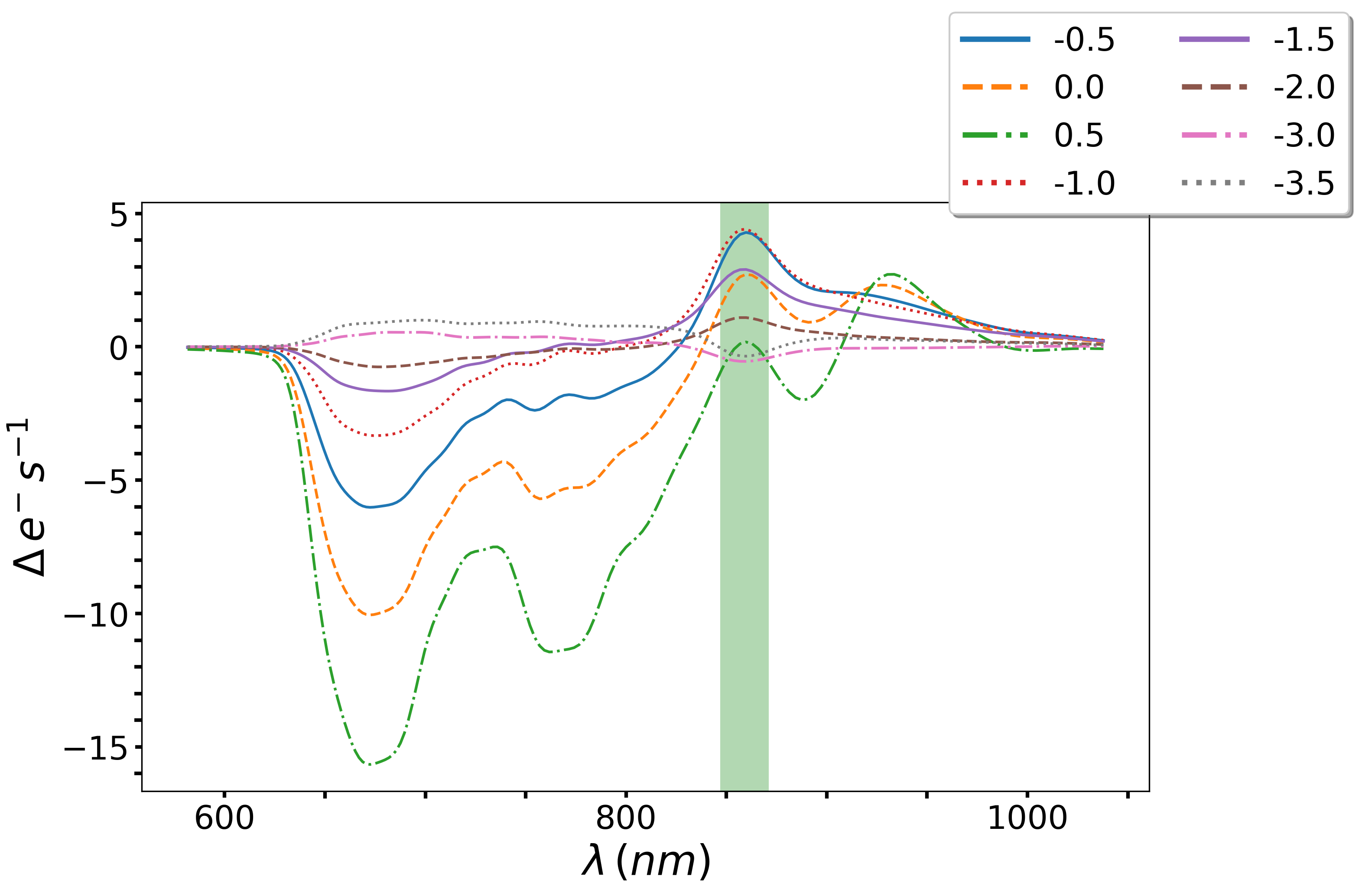}
    \caption{Differential BP (top panel) and RP (bottom panel) spectra of Solar $\mathrm{[C/Fe]}$-scaled stars, with different $\mathrm{[Fe/H]}$. Each differential spectrum results from $reference \, spectrum - spectrum,$ where $\mathrm{[Fe/H]_{ref}=-2.5}$. The shaded areas correspond to the regions we used for our flux ratios, i.e., the Ca II H \& K region (purple), the G-band (magenta), the H$\beta$ (yellow), and the Ca II NIR (green) regions. }
   \label{fig:diff_spec1}
\end{figure}

Differential simulated spectra were used to find out how the flux changes with changing metallicity. For this exercise, we used the spectra with $\mathrm{[Fe/H]=-2.5\,dex}$ as a reference, and we subtracted the spectra of lower and higher metallicity that had all the other astrophysical parameters identical to the reference spectrum (Figure \ref{fig:diff_spec1}). The regions we expected to be $\mathrm{[Fe/H]}$ sensitive were the Ca II H \& K and G-band wavelength ranges, which was confirmed: the lower the metallicity of the object, the higher the flux. The regions of the spectra that cover the Ca II near-infrared triplet as well as the H$\beta$ absorption line showed a greater sensitivity to the change in all the other stellar parameters compared to $\mathrm{[Fe/H]}$. Using the ratios of the aforementioned spectral regions, that is the ratio of the integrated Ca H \& K flux to that of the H$\beta$ region ($fr\mathrm{_{CaHK/H\beta}}$), and that of the G band integrated flux to the Ca NIR triplet ($fr\mathrm{_{G/CaNIR}}$), we can see a relation with metallicity (Figure \ref{fig:met_relation}). Figure \ref{fig:met_relation} was created from our entire dataset, that is temperature, surface gravity (Figure \ref{fig:isochrone_params}), and all physically meaningful $\mathrm{[C/Fe]}$-$\mathrm{[Fe/H]
}$ combinations, in  other words $\mathrm{[C/Fe]+[Fe/H]\leq -1\,dex}$ for $\mathrm{[Fe/H]}\leq0.5$ and $-1.5\geq\mathrm{[C/Fe]\leq0}$ for $\mathrm{[Fe/H]}\geq-0.5$.

\subsection{Metallicity-temperature-$fr\mathrm{_{CaHK/H\beta}}$ relation}\label{sec_mtfr}
\begin{figure}
   \centering
   \includegraphics[width=0.5\textwidth]{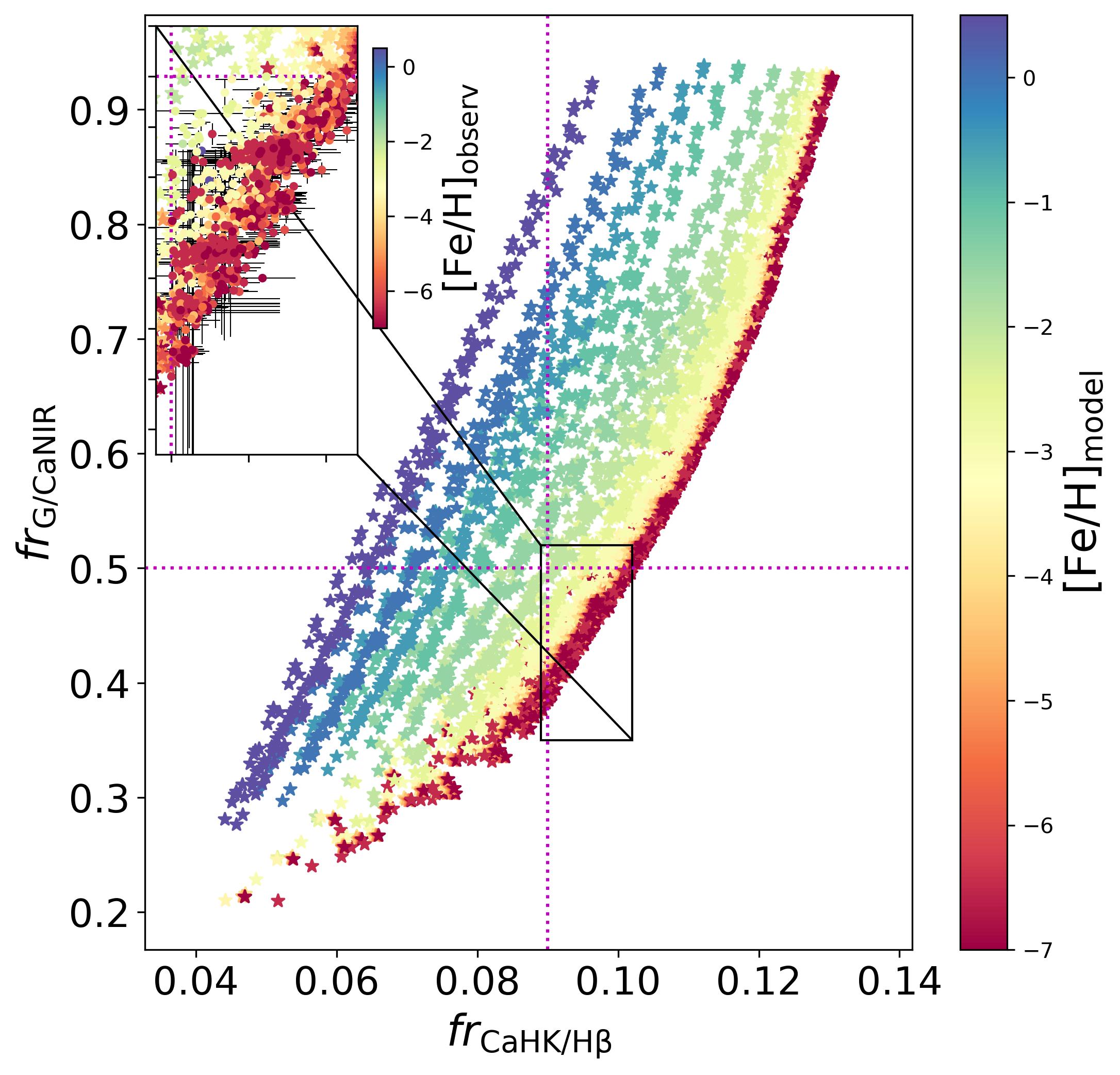}
    \caption{Smoothly changing metallicity for model stars of G=15 mag. On the  inset plot, it is shown how well the noisy simulated flux ratios track the modeled (noiseless) ones.}
   \label{fig:met_relation}
\end{figure}

\begin{figure}
   \centering
   \includegraphics[width=0.5\textwidth]{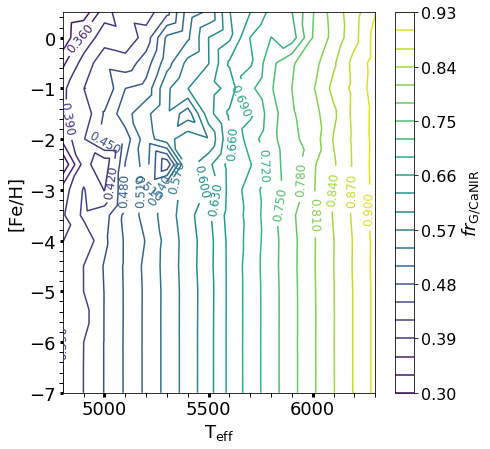}
    \caption{$fr\mathrm{_{G/CaNIR}}$ is an effective temperature indicator; however, it does not carry information regarding the iron abundance.}
   \label{fig:teff_proxy}
\end{figure}

\begin{figure}
   \centering
   \includegraphics[width=0.5\textwidth]{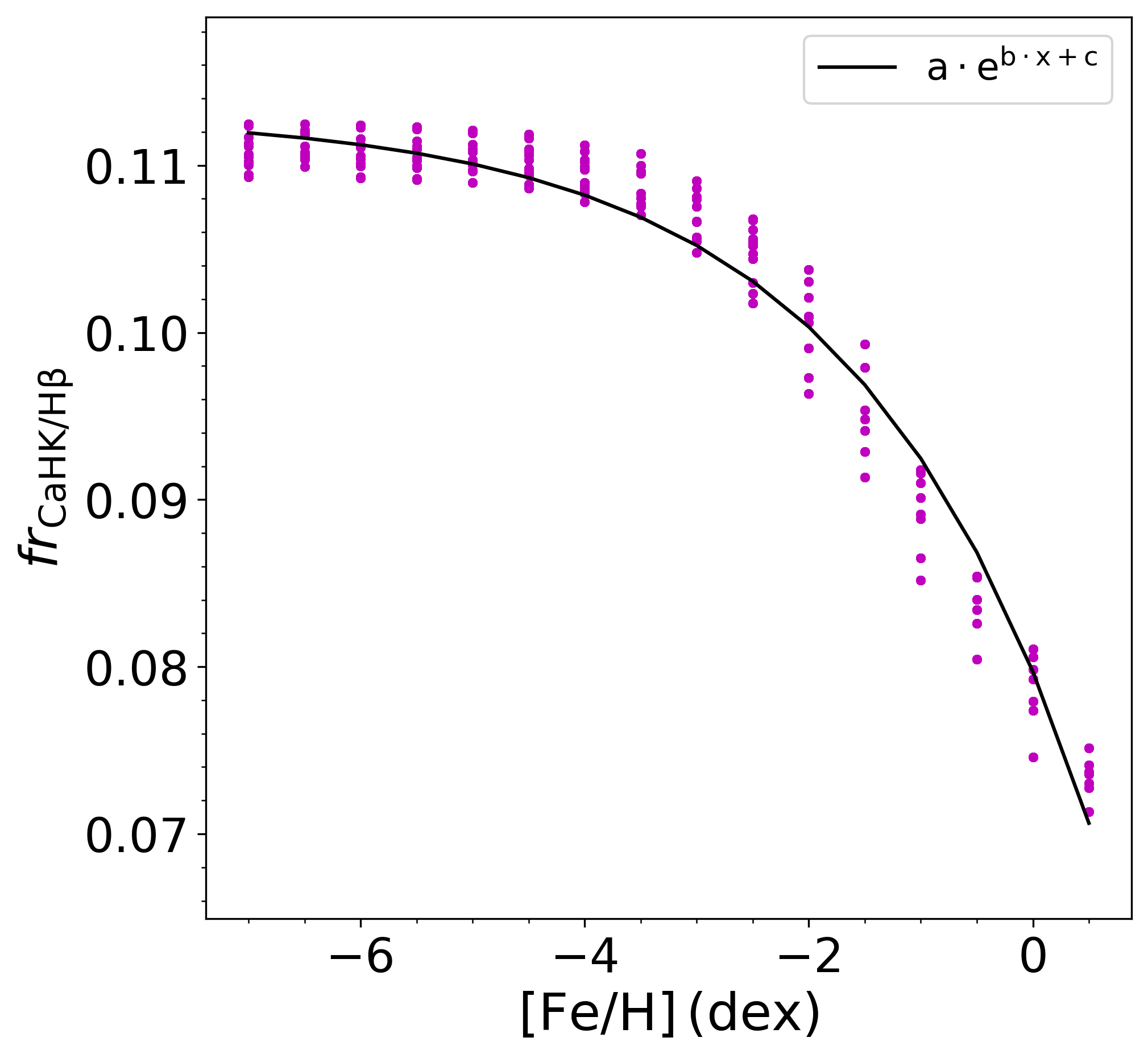}
    \caption{Exponential decline of $fr\mathrm{_{CaHK/H\beta}}$ with increasing [Fe/H], for a roughly constant $fr\mathrm{_{G/CaNIR}}$. This behavior starts to break down for $\mathrm{T_{eff}<4800\,K}$. The scatter at each iron abundance reflects the different carbon enhancement and surface gravity, as well as a moderate ($\sim 300$K) variation in temperature.}
   \label{fig:constant_Faa}
\end{figure}

The relation between $fr\mathrm{_{CaHK/H\beta}}$, $fr\mathrm{_{G/CaNIR}}$, and $\mathrm{[Fe/H]}$ that is emerging in Figure \ref{fig:met_relation} encouraged us to try to find a relation of the form $F(fr\mathrm{_{CaHK/H\beta}},fr\mathrm{_{G/CaNIR}})=\mathrm{[Fe/H]}$. In this venture, we found that $fr\mathrm{_{G/CaNIR}}$ is -- by itself -- an indicator for the effective temperature (see Figure \ref{fig:teff_proxy}), but it does not carry any information about the metallicity. On the other hand, for constant $fr\mathrm{_{G/CaNIR}}$, the metallicity is changing smoothly with the change of $fr\mathrm{_{CaHK/H\beta}}$ (Figure \ref{fig:constant_Faa}), hence $fr\mathrm{_{CaHK/H\beta}}$ carries information concerning the iron abundance. There is, however, a scatter that accounts for variations in $\mathrm{T_{eff}}\, ,\mathrm{\log g}$, and $\mathrm{[C/Fe]}$. So, since $fr\mathrm{_{G/CaNIR}}$ is not metallicity-sensitive, as mentioned above, but rather a temperature indicator, we thought it best to proceed with the objective of developing a relation which is comprised of $fr\mathrm{_{CaHK/H\beta}}$ and some relevant stellar parameters. The usage of stellar parameters makes the range of applicability more straightforward to implement, and subsequently more user-friendly. 
 By inspecting Figure \ref{fig:constant_Faa}, we assessed that the therein $fr\mathrm{_{CaHK/H\beta}}$-$\mathrm{[Fe/H]}$ correlation can be described with an exponential relation (Figures \ref{fig:met_relation} and \ref{fig:constant_Faa}), which was confirmed by the residuals of the fit. This exponential relation can then be expressed with respect to the temperature and surface gravity of the model stars (Figure \ref{fig:classification}), enabling us to use them as priors, for example from Gaia itself \citep{2018A&A...616A...8A}. Finally, we can use the iso-$fr\mathrm{_{CaHK/H\beta}}$ lines (see Figure \ref{fig:classification}), as well as the effective temperatures and surface gravities, in order to infer the metallicity:

\begin{equation}
 F(fr\mathrm{_{CaHK/H\beta}},\mathrm{T_{eff}},\mathrm{\log g})=\mathrm{[Fe/H]}\label{eq:1}
\end{equation}

\begin{gather*}
F=-(\mathrm{T_{eff}}\cdot \mathrm{\log g}) \cdot e^{b\cdot fr\mathrm{_{CaHK/H\beta}} + c}+d, \tag{2} \label{eq:2}\\
\end{gather*}
where  $b,c,$ and $d$ are $\mathrm{T_{eff}}$ and $\mathrm{\log g}$ dependent coefficients.

The coefficients $b,c,d$ are the result of fitting Eq. \ref{eq:2} to $fr\mathrm{_{CaHK/H\beta}}$-metallicity pairs of a roughly constant temperature -- $\mathrm{T_{eff}}\mp30$K -- and constant surface gravity (see Figure \ref{fig:mfunction_fitting}) for metallicities greater than $-3.5$ dex. The aforementioned $fr\mathrm{_{CaHK/H\beta}}$-$\mathrm{[Fe/H]}$ pairs result from the iso-$fr\mathrm{_{CaHK/H\beta}}$ lines.

   
\section{Results}\label{sec_results}
We first tested our method on the very same data we used to construct it, and then we examined how it is influenced by different stellar parameters. Finally, we applied our method to noisy spectra.  \newline
\subsection{Testing on the model spectra}
\begin{figure}
   \centering
   \includegraphics[width=0.5\textwidth]{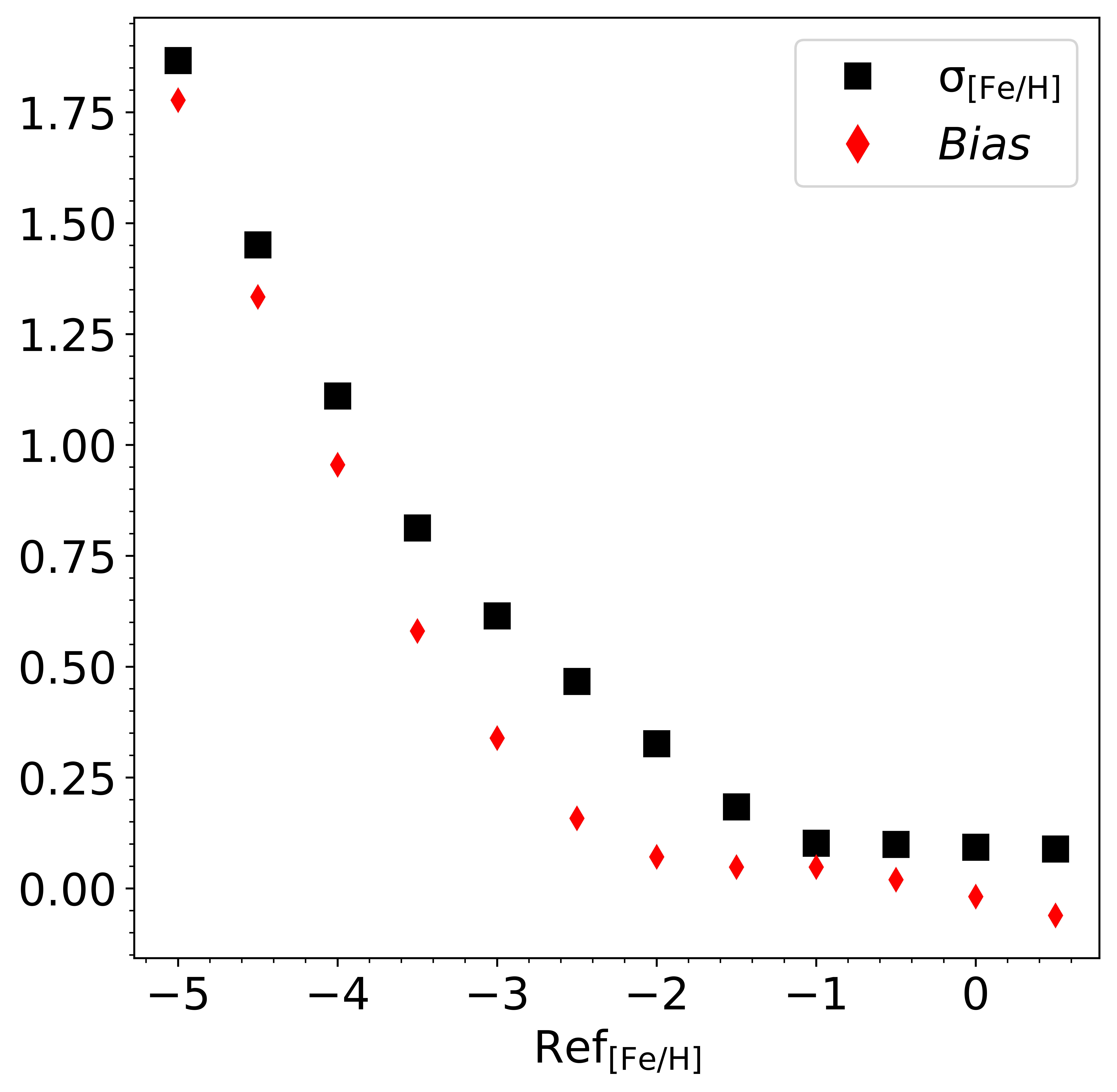}
    \caption{$\sigma$ and bias of the inferred metallicities of our noise-free spectra of G=15 mag.}
   \label{fig:mfunction_sigma_15mag_model}
\end{figure}

We applied our method on the same model spectra we used to find relation \ref{eq:2}. Figure \ref{fig:mfunction_sigma_15mag_model} shows that our method works very well for stars with $\mathrm{[Fe/H]}\geq-3$ dex since $\sigma_\mathrm{{[Fe/H]}}\leq0.6$ dex. Specifically, for red giant branch stars $\sigma_\mathrm{{[Fe/H]}}=0.6$ and for turnoff stars, $\sigma_\mathrm{{[Fe/H]}}\approx0.3$ at $\mathrm{[Fe/H]}=-3$. Further, even for stars with $\mathrm{[Fe/H]}=-3.5$ dex, $\sigma$ is smaller than $1.0$ dex. On the other hand, we can identify stars with $\mathrm{[Fe/H]}\geq-1$ very accurately, that is to say with $\sigma_\mathrm{{[Fe/H]}}\approx0.1$ dex. The bias of these results (see Figure \ref{fig:mfunction_sigma_15mag_model}), which we defined as $\overline{\mathrm{[Fe/H]_{inf}}-\mathrm{[Fe/H]_{ref}}}$, shows that our method tends to overestimate the metallicity for $\mathrm{[Fe/H]_{ref}}\leq-1$, in other words it assigns higher values than the true ones. The benefit of this result is that the inferred metallicities that are in and below the metal-poor range are most probably as low as or even lower than the true ones (see Section \ref{sec_sucrate}). Further, the way the difference between the bias and $\sigma$ decreases as we go to lower $\mathrm{[Fe/H]_{ref}}$ indicates that there is a point below which we cannot distinguish metallicities, and that is around $\mathrm{[Fe/H]_{ref}}=-3.5$ dex.
\subsection{Application on noisy spectra and the dependence on stellar parameters}\label{sec_appl}
\begin{figure*}
   \centering
   \includegraphics[width=0.7\textwidth]{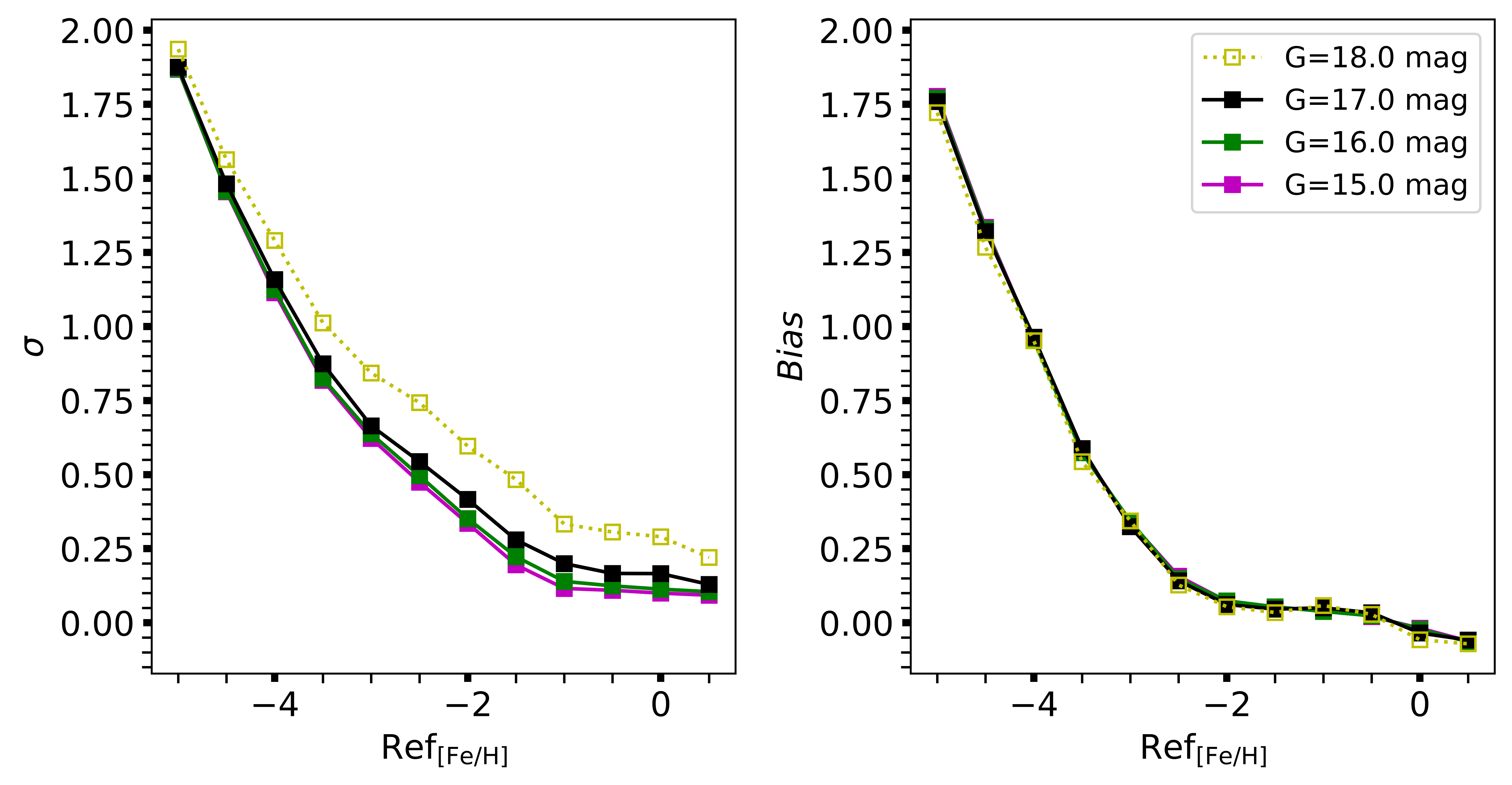}
    \caption{$\sigma$ and bias of the inferred metallicities of noisy spectra of different magnitudes: G=15, 16, 17, 18 mag.}
   \label{fig:mfunction_sigma_all}
\end{figure*}

\begin{figure}
   \centering
   \includegraphics[width=0.3\textwidth]{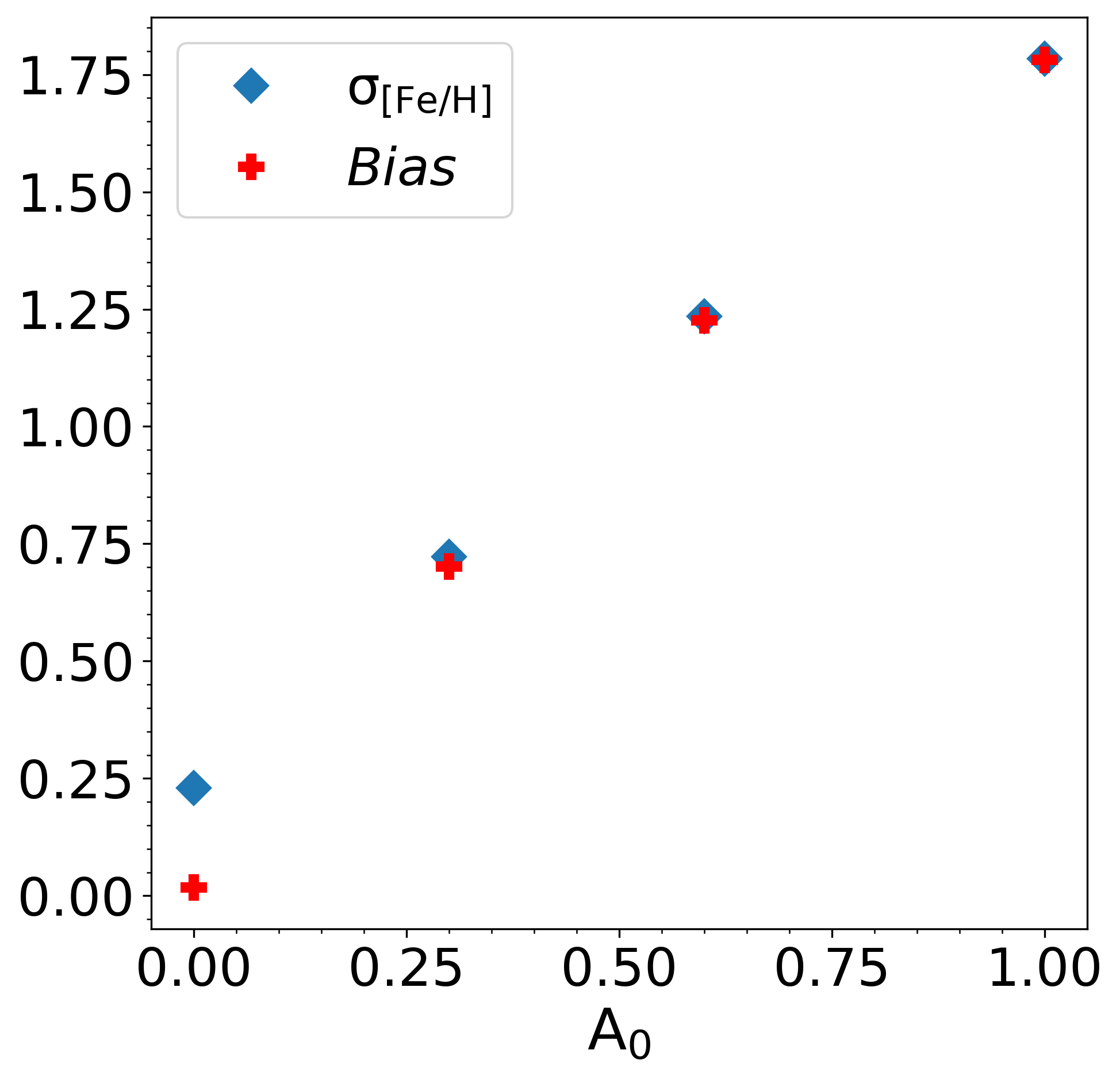}
    \includegraphics[width=0.3\textwidth]{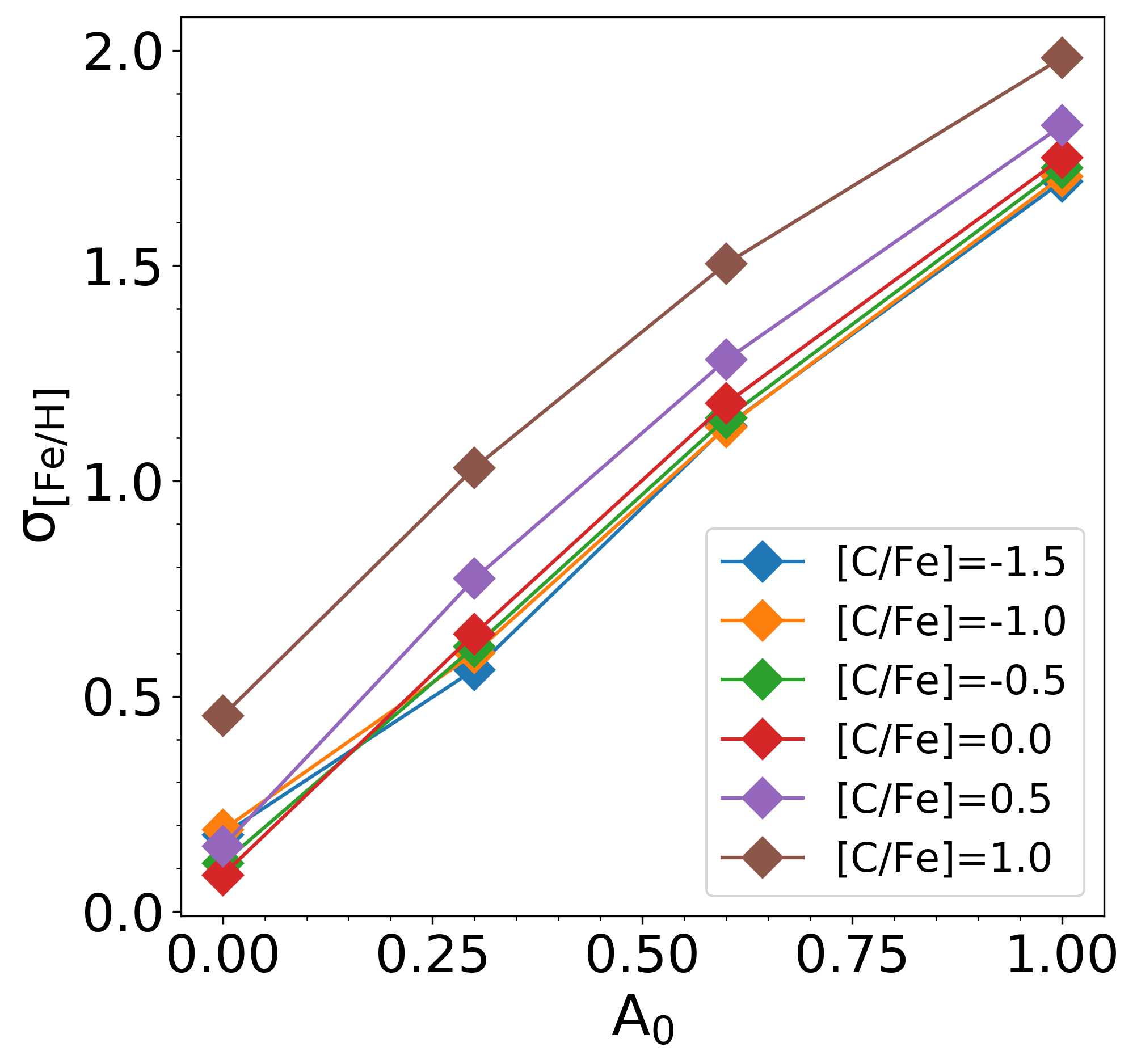}
    \caption{Testing the dependence of metallicity inference on different extinction coefficients has the following expected outcome: the higher the extinction, the higher the uncertainty. We used one set of stellar parameters and metallicity, i.e., $\mathrm{T_{eff}=5500}$ K, $\mathrm{\log g=3.5}$ dex, and  $\mathrm{[Fe/H]=-2}$ dex, but various relative carbon abundances. On the top are the $\sigma_\mathrm{{[Fe/H]}}$ and bias in bins of extinctions, and in the bottom the results are also in bins of different $\mathrm{[C/Fe]}$.}
   \label{fig:mfunction_A0_test}
\end{figure}

 We subsequently applied our method to noisy spectra. 
 For each of our model spectra (G=15mag), we generated 20 noisy ones: we inferred the metallicity with $\sigma_\mathrm{{[Fe/H]}}\leq0.6$ dex for $\mathrm{{[Fe/H]}}\geq-3$ dex (Figure \ref{fig:mfunction_sigma_all}). Again, the uncertainty in inferring metallicity for metal-poor stars ($\mathrm{[Fe/H]}=-1$) and above is very low with $\sigma_\mathrm{{[Fe/H]}}\lessapprox0.12$ dex.
 Further we sought to determine how our method performs in the different temperature bins. For this purpose, we computed $\sigma_\mathrm{{[Fe/H]}}$ for all spectra of the same temperature, surface gravity, and relative carbon abundance (Figure \ref{fig:mfunction_sigma_teff_15mag}). We found that the dependence on temperature is linked with $\mathrm{[C/Fe]}$. When $\mathrm{[C/Fe]}\leq 0.0$ dex, our method performs slightly better for lower effective temperatures of the same $\mathrm{\log g}$ when $\mathrm{{[Fe/H]}<-4}$. In contrast, when the iron abundance is greater than -4,  performance becomes independent of $\mathrm{T_{eff}}$. When $\mathrm{[C/Fe]>0}$, performance has a similar behavior for $\mathrm{{[Fe/H]}<-4}$, that is to say it is better for lower temperatures.\ However, above that threshold, the effect is reversed: the performance is much better for higher effective temperatures. That phenomenon is intensified as $\mathrm{[C/Fe]}$ grows. We expected that we could determine the metallicity for lower temperatures more accurately since the Ca H \& K lines are stronger in cooler stars compared to hotter ones, even down to very low metallicities. This, as stated above, is the case for $\mathrm{[C/Fe]\leq0}$, but not above that threshold. The reason for this is probably due to the many carbon lines that arise as a result of the high carbon enhancements, which in turn are also stronger at lower effective temperatures and, consequently, our method assigns higher metallicities to those spectra. Furthermore, we investigated the dependence on different relative carbon to iron abundances, for which we already have some information as stated above. To test that, we separated our data into surface gravity and $\mathrm{[C/Fe]}$ bins, and calculated the error of the inferred metallicities (Figure \ref{fig:mfunction_c_impact_15mag}). There is an obvious difference in the performance pattern for stars below and above $\mathrm{[C/Fe]}=0$ dex. At and below that threshold, $\sigma_\mathrm{[Fe/H]}$ is almost independent of $\mathrm{[C/Fe]}$. Above it, however, the performance declines as the relative carbon abundance is rising. This effect though is attenuated as temperature rises.\newline
 Another factor we considered is the extinction $A_{0}$. We used synthetic spectra with $\mathrm{T_{eff}=5500}$ K, $\mathrm{\log g=3.5}$ dex, $\mathrm{[Fe/H]=-2}$ dex, and all possible $\mathrm{[C/Fe]}$ combinations -- for which $\mathrm{[Fe/H]+[C/Fe]}\geq-1$ is true -- and passed them through Ulysses generating spectra of two kinds for each model:  a noiseless spectrum and 20 noisy ones. We expected that the performance would decline with greater extinction, which is what we observed (see Figure \ref{fig:mfunction_A0_test}).  We validated this result by performing the above exercise for a collection of spectra with varying astrophysical parameters. As extinction rises, $fr\mathrm{_{CaHK/H\beta}}$ and $fr\mathrm{_{G/CaNIR}}$ decrease, in other words the model star appears to be colder and the Ca H \& K features are less distinguishable. When we use Figure \ref{fig:met_relation} as a reference, a model star with increasing extinction shifts left and downwards in the plot. However, according to \cite{2008A&A...484..721C}, about $90\%$ of stars at high galactic latitudes (i.e., at $|b|\geq20^o$) have a reddening that is smaller than 0.06 ($E(B-V)\leq0.06$), which corresponds to $A_v<0.19$. Thus, the issue of extinction is outside the scope of this paper, but we consider addressing its effect in our follow-up work. \newline
  Lastly, we wanted to find out for which magnitude our method starts to break down. Hence, we generated noisy spectra (one noisy spectrum for each set of stellar parameter combinations) of magnitudes G=16, 17, 18 mag. Up to 17 mag, the performance of our method is roughly on the same levels of precision (Figure \ref{fig:mfunction_sigma_all}), and at 18 mag, $\sigma_\mathrm{[Fe/H]}]<0.85$ for $\mathrm{[Fe/H]\geq-3}$. Our method does not seem to break down, but it rather gets less precise as the magnitude rises. Even so, at 18 mag, we can infer metallicities down to $-2$ dex with an uncertainty of $\sigma_\mathrm{[Fe/H]}]\leq0.6$ dex.

\subsection{Predicted success rate}\label{sec_sucrate}
\begin{figure}
   \centering
   \includegraphics[width=0.5\textwidth]{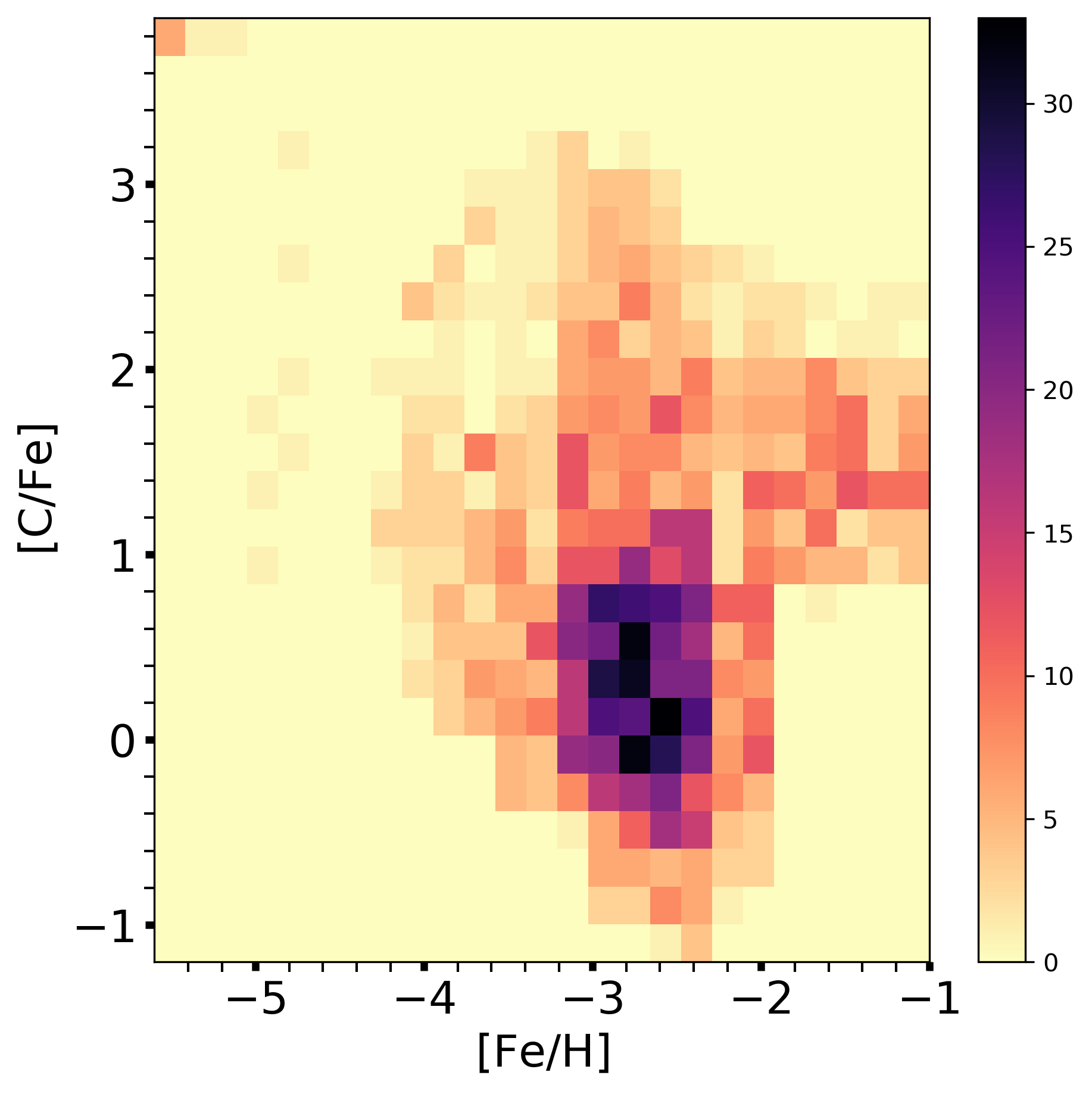}
    \caption{Metallicity and carbonicity of the simulated dataset we used to predict the success rate of our method. The color bar designates the number of models in each $\mathrm{[Fe/H]}$-$\mathrm{[C/Fe]}$ bin.}
   \label{fig:MCL_test}
\end{figure}

\begin{figure}
   \centering
   \includegraphics[width=0.5\textwidth]{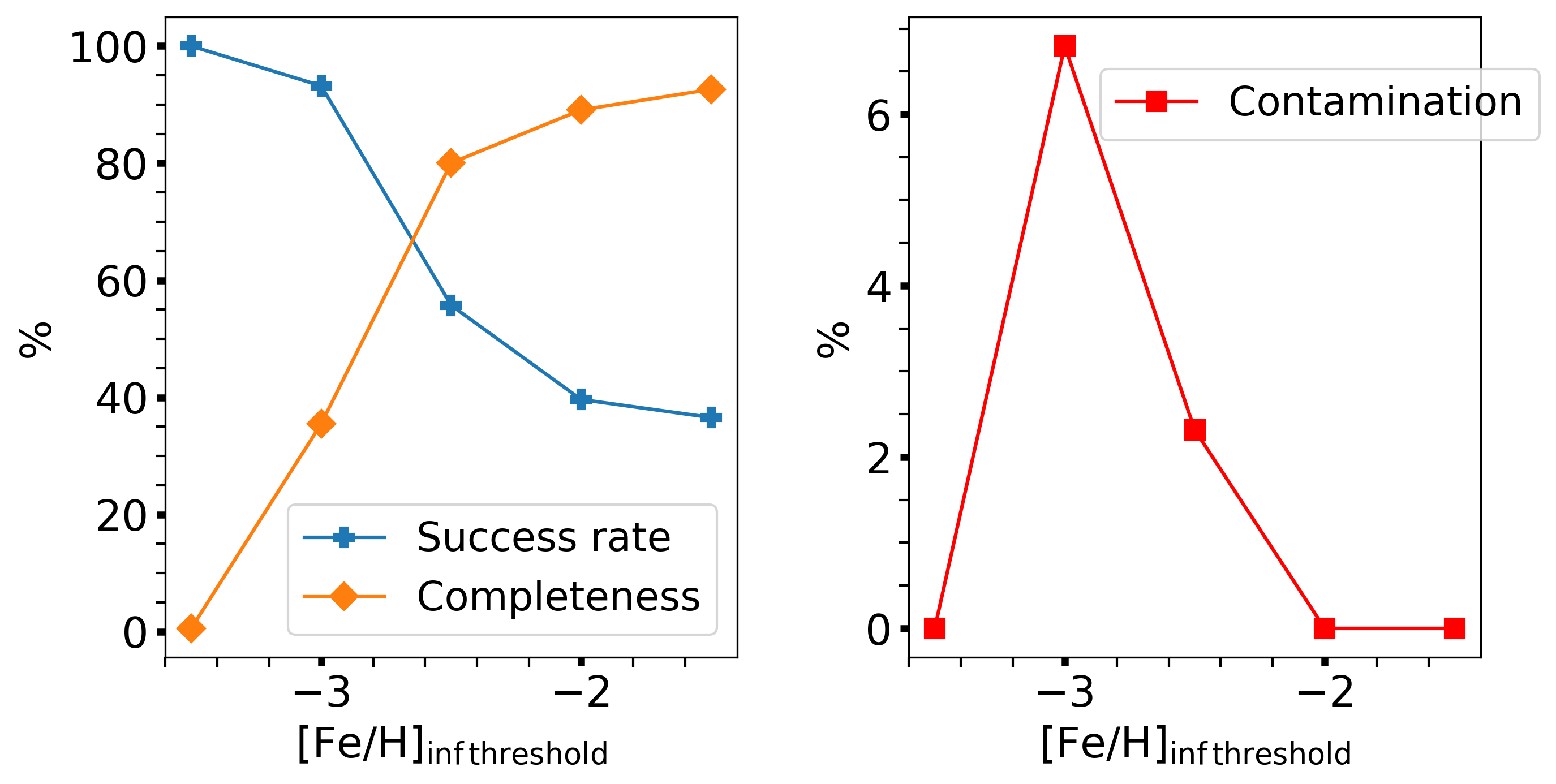}
    \caption{Calculation of the success rate, the completeness (left plot), and the contamination (right plot) for threshold values $\mathrm{[Fe/H]_{inf}}=-3.5,-3.0,-2.5,-2.0$. The contamination was calculated with reference to the threshold metallicity, whereas the completeness and the success rate were calculated with reference to model spectra with $\mathrm{[Fe/H]}\leq-3$.} 
   \label{fig:MCL_stats}
\end{figure}

Lastly, an application of our method to a simulated dataset (G=15 mag) of realistic $\mathrm{[Fe/H]}$ and $\mathrm{[C/Fe]}$ distributions (Figure \ref{fig:MCL_test}) served the purpose of assessing the expected success rate of our method. Additionally, this test enabled us to set the metal-poor threshold, that is to say the derived $\mathrm{[Fe/H]}$ value below which metal-poor stars can be selected, with the greatest efficiency and completeness. We constructed the aforementioned dataset in the following manner: a) we randomly drew $\mathrm{T_{eff}}$-$\mathrm{\log g}$-$\mathrm{[Fe/H]}$-$\mathrm{[C/Fe]}$ combinations from our parameter space; b) then we replaced the metallicity and carbonicity values by drawing new ones from two different metallicity distribution functions (MDF) of halo stars, and from two carbonicity distribution functions, respectively. Specifically, we used the MDF from \cite{2020MNRAS.492.4986Y} when the initially drawn metallicity was $>-2$ dex, and the MDF from \cite{2014ApJ...797...21P} otherwise. Concerning the carbonicity, we used the respective carbon-enhancement distribution for the \cite{2014ApJ...797...21P} MDF \citep{2014ApJ...797...21P}, and we used the $\mathrm{[C/Fe]}$ distribution from \cite{2016ApJ...833...20Y} for $\mathrm{[Fe/H]}>-2$. \newline We found that when we selected all stars with an inferred $\mathrm{[Fe/H]_{inf}}\leq-2.5$, we recovered $80\%$ of stars with $\mathrm{[Fe/H]}\leq-3$, and we had a contamination of about 2\% of stars with metallicities above -2.5. It should be noted, however, that these ``contamination" stars all have  $\mathrm{[Fe/H]}<-2$. Furthermore, about 55\% of the stars with $\mathrm{[Fe/H]_{inf}}\leq-2.5$ have a reference $\mathrm{[Fe/H]}\leq-3$, which means that one in two of the selected stars would at least be extremely metal-poor ($\mathrm{[Fe/H]}\leq-3$). If we were to select the metallicity threshold at $\mathrm{[Fe/H]_{inf}}\leq-3$, the hit rate for stars below -3 would increase to 9 out of 10. We would, however, fail to detect about 75\% of stars with $\mathrm{[Fe/H]}\leq-3$ (see Figure \ref{fig:MCL_stats}). The overall results of this test are detailed in Figure \ref{fig:MCL_stats}. Specifically, we show how the success rate, the completeness, and the contamination change by selecting a different threshold, that is for $\mathrm{[Fe/H]_{inf\,threshold}}=-2.0,-2.5,-3.0,-3.5$ dex. We define the success rate as the percent of the selected stars that actually have $\mathrm{[Fe/H]_{ref}}\leq-3.0$, the completeness as the percent of the total number of stars with $\mathrm{[Fe/H]_{ref}}\leq-3.0$ that have $\mathrm{[Fe/H]_{inf}}\leq\mathrm{[Fe/H]_{inf\,threshold}}$, and the contamination as the percent of selected stars that have $\mathrm{[Fe/H]_{ref}}>\mathrm{[Fe/H]_{inf}}$.

\section{Conclusions}

      We developed a method using flux ratios of metallicity-sensitive regions from the Gaia BP/RP low resolution spectra in order to find new metal-poor stars.  This method is applicable when stars have $\mathrm{4800\,K \geq T_{eff}\leq6300\,K}$. We took into account the fact that a large fraction of metal-poor stars are carbon enhanced, and thus used a grid of synthetic spectra with varying $\mathrm{[Fe/H]}$ and $\mathrm{[C/Fe]}$. We found an exponential relation between the metallicity and the $fr\mathrm{_{CaHK/H\beta}}$ flux ratio, which is temperature and surface gravity dependent. Therefore, our method requires both of those stellar parameters as priors. We first applied our method to the very same noiseless data ($\mathrm{G=15\,mag}$) we used to construct it, and $\mathrm{[Fe/H]}$ was inferred with an uncertainty of $\sigma\lessapprox0.6$ dex at $\mathrm{[Fe/H]}\leq-3.0$ dex. Our method's performance was approximately on the same level for noisy spectra of the same and greater magnitudes, that is $\sigma\lessapprox0.65$ for $\mathrm{[Fe/H]}\leq-3.0$ dex and $\mathrm{G=15,16,17\, mag}$. We found that the performance depends on temperature, but in conjunction with the relative carbon abundance: for Solar $\mathrm{[C/Fe]}$ and below, performance is slightly better for lower temperatures of the same surface gravity when determining $\mathrm{[Fe/H]}\lessapprox-3.5$. For $\mathrm{[Fe/H]}\geq-3$, the performance is practically independent of $\mathrm{T_{eff}}$. When $\mathrm{[C/Fe]}>0$, $\sigma_{\mathrm{[Fe/H]}\lessapprox-3.5}$ is as in the Solar case, that is it is lower for lower $\mathrm{T_{eff}}$. However, the determination of the metallicity above $-3.5$ dex presents a lower uncertainty at higher temperatures of the same $\mathrm{\log g}$. A dependence of the performance from the relative carbon-to-iron abundance is observed when $\mathrm{[C/Fe]}>0$. Further, even for spectra of $\mathrm{18^{th}}$ magnitude, we can determine metallicities down to $-2$ with an uncertainty of $\sigma_{\mathrm{[Fe/H]}}<0.6$ dex, which is sufficient for identifying metal-poor stars. Nevertheless, we find that the overall performance of our method deteriorates with rising extinction $A_{0}$. Lastly, we predict that by selecting stars with $\mathrm{[Fe/H]_{inf}}\leq-2.5$ dex, we achieve a completeness of 80\% of the stars with $\mathrm{[Fe/H]}\leq-3$, and that one in two stars we select is extremely metal-poor. Our method will be applied to Gaia DR3 data and the results will be published in a forthcoming paper.

\begin{acknowledgements}
    This work was funded by the Deutsche Forschungsgemeinschaft (DFG, German Research Foundation) -- Project-ID 138713538 -- SFB 881 (``The Milky Way System'', subproject A04). 
    KL acknowledges funds from the European Research Council (ERC) under the European Union’s Horizon 2020 research and innovation program (Grant agreement No. 852977) and funds from the Knut and Alice Wallenberg foundation. This research was supported by the Australian Research Council Centre of Excellence for All Sky Astrophysics in 3 Dimensions (ASTRO 3D), through project number CE170100013. This work was supported by computational resources provided by the Australian Government through the National Computational Infrastructure (NCI) under the National Computational Merit Allocation Scheme and the ANU Merit Allocation Scheme (project y89).
\end{acknowledgements}

\bibliography{references}
    \bibliographystyle{aa}
    
\begin{appendix}
\onecolumn
\section{Additional figures and data.}
\begin{figure*}[b!]
   \centering
   \includegraphics[width=0.7\hsize]{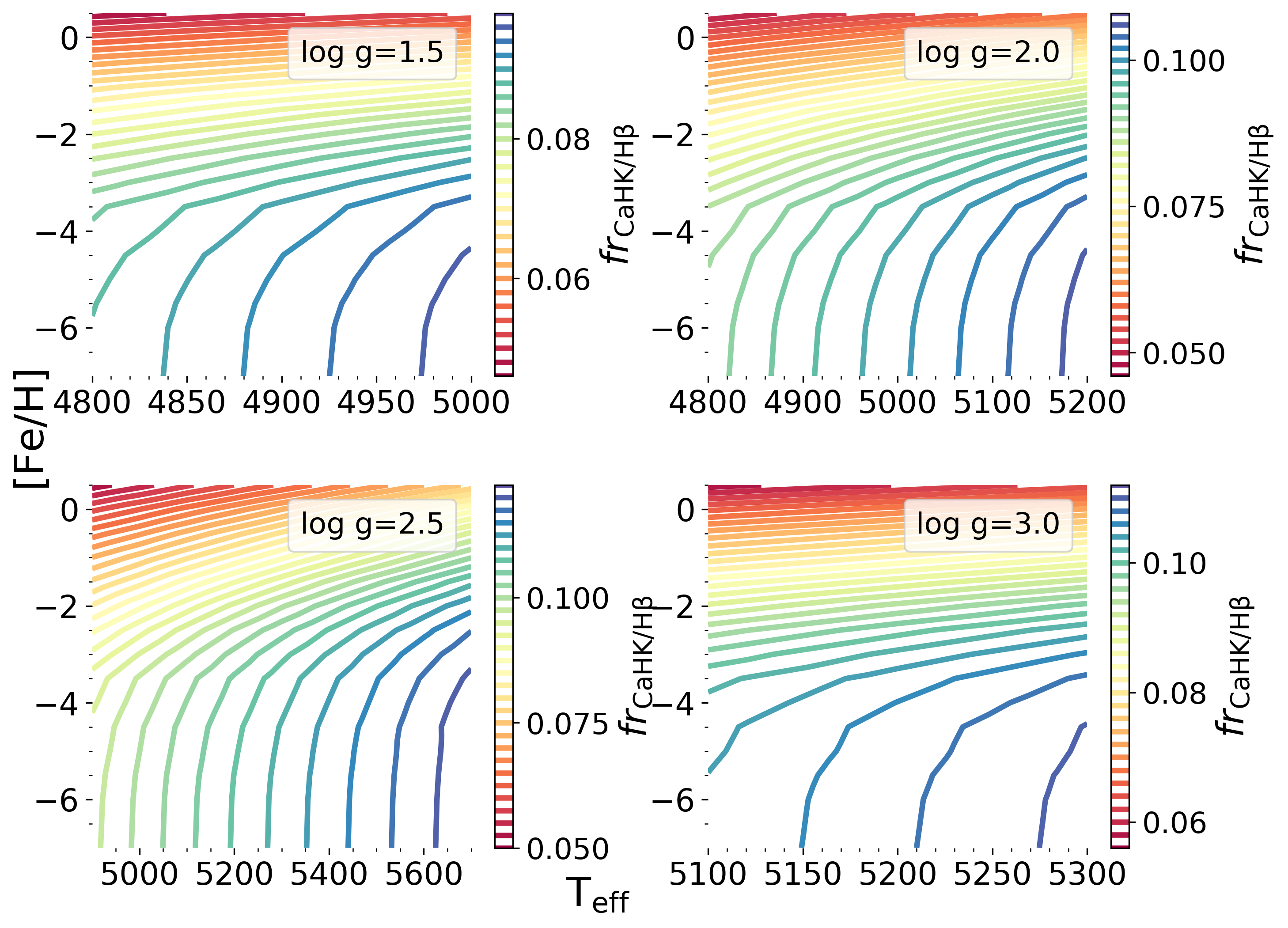}
   \includegraphics[width=0.7\hsize]{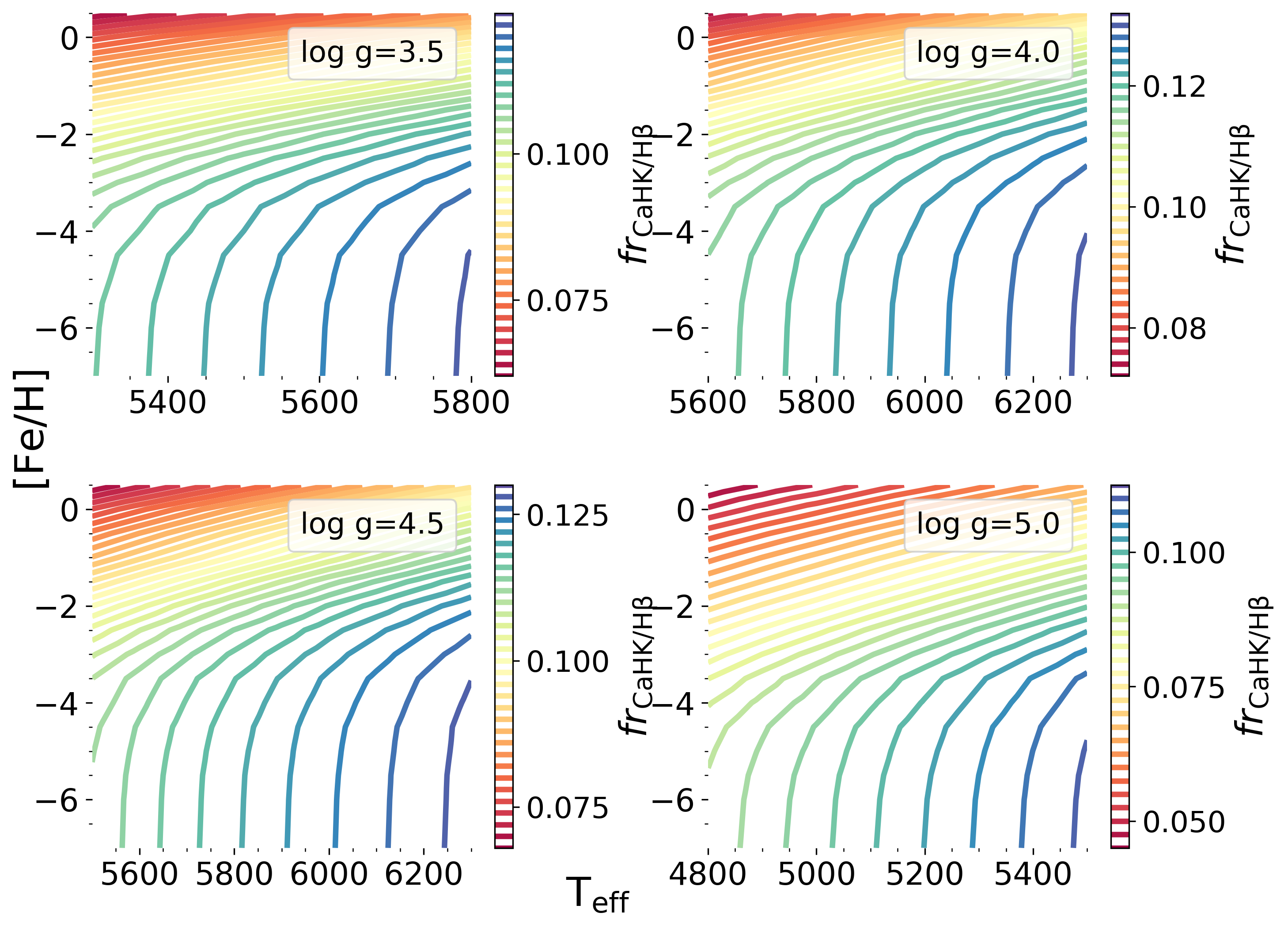}
    \caption{Contour plots where the third dimension are iso-flux lines, in particular for the $fr\mathrm{_{CaHK/H\beta}}$ flux ratio. Those iso-$fr\mathrm{_{CaHK/H\beta}}$ lines are temperature-sensitive up to about  $\mathrm{[Fe/H}]=-3.5$ dex, and then they become  metallicity-sensitive, hence we can use them to find a relation from which we can infer $\mathrm{[Fe/H]}$. Because the metallicity sensitivity starts at  $\mathrm{[Fe/H}]=-3.5$, we cannot distinguish iron abundances below that threshold, but rather identify whether they are below or above it.}
   \label{fig:classification}
\end{figure*}

\begin{figure*}
   \centering
   \includegraphics[width=\textwidth]{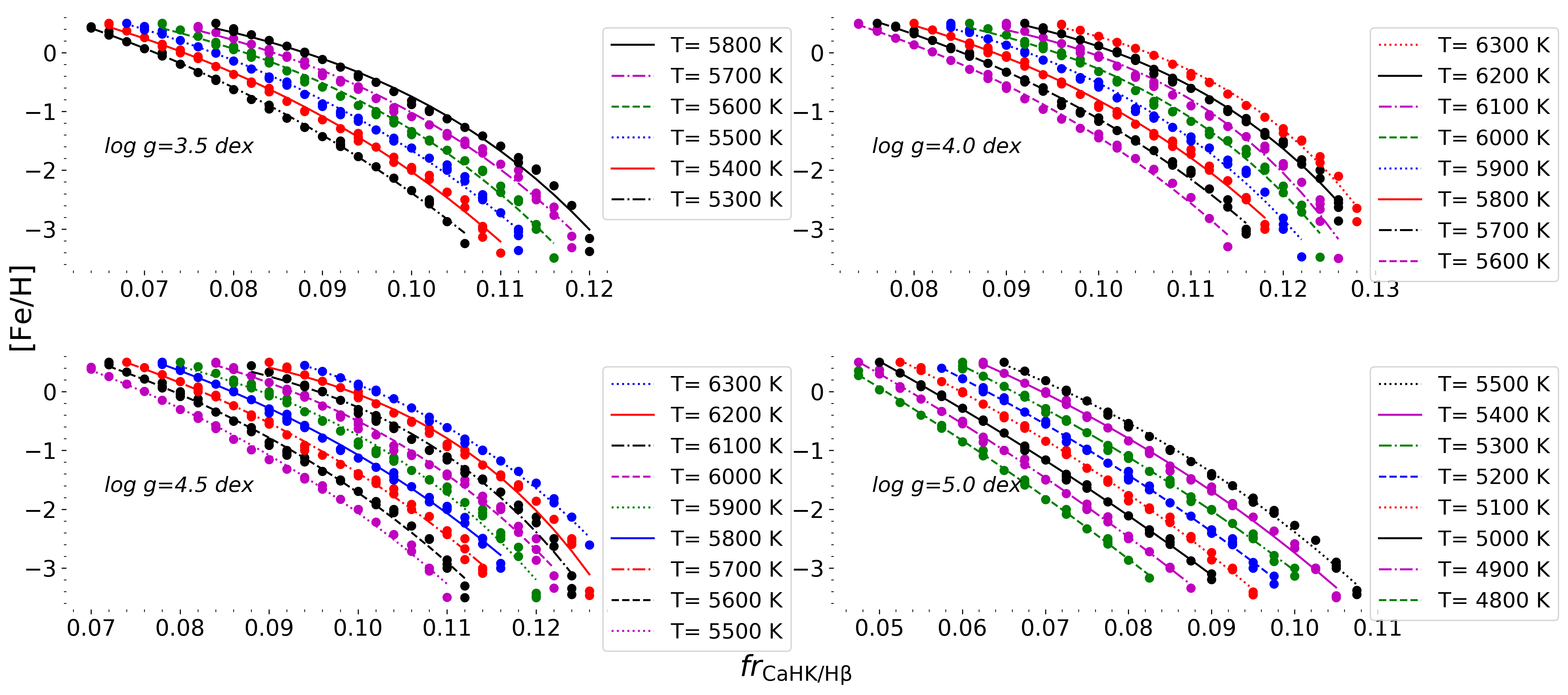}
    \caption{We fit Equation \ref{eq:2} in  the metallicity-$fr\mathrm{_{CaHK/H\beta}}$ space for each temperature-surface gravity pair from our parameter space (see Figure \ref{fig:isochrone_params}). Some of those fits are shown in the above panels.}
   \label{fig:mfunction_fitting}
\end{figure*}

\begin{figure}
   \centering
   \includegraphics[width=0.5\textwidth]{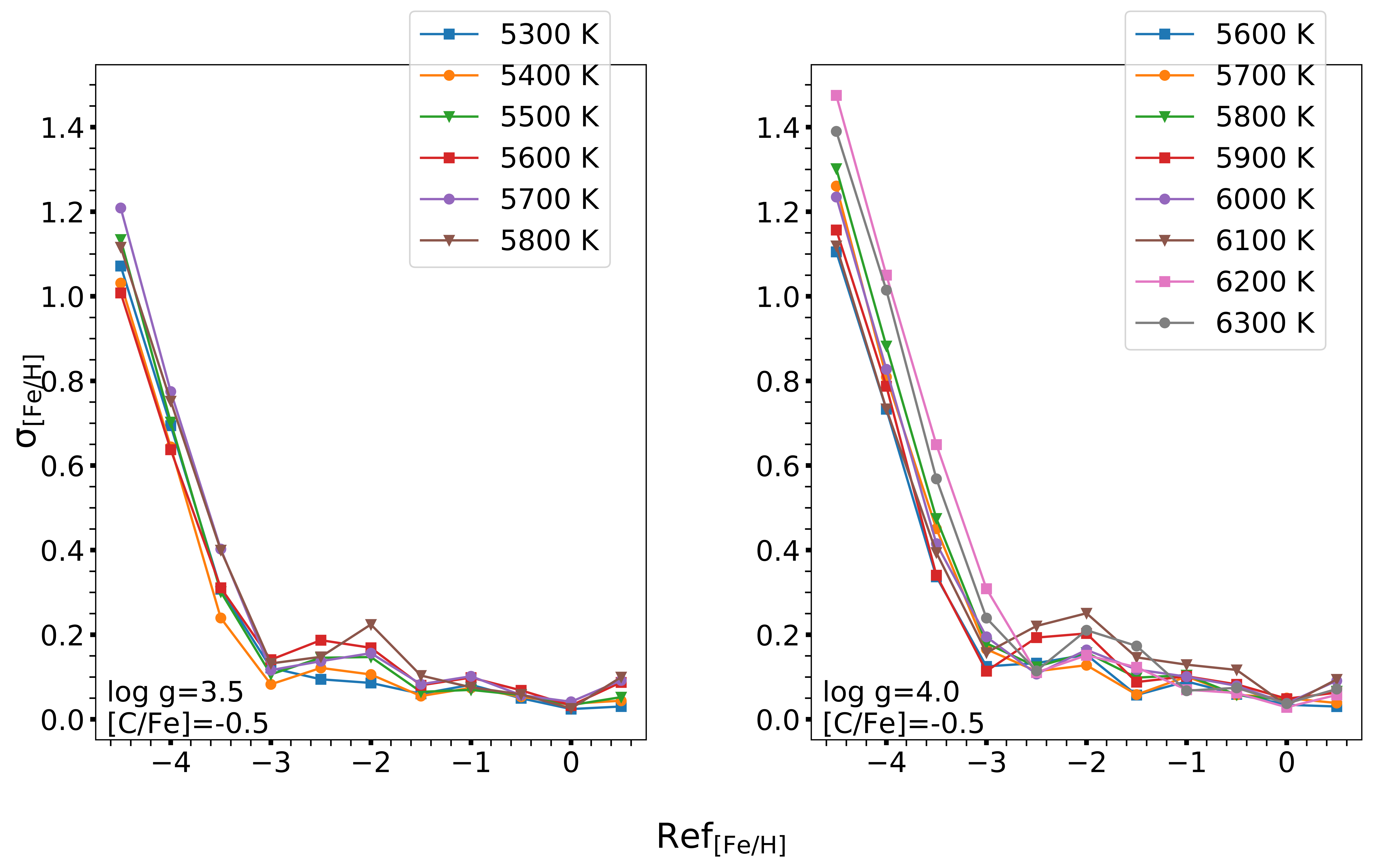}
    \includegraphics[width=0.5\textwidth]{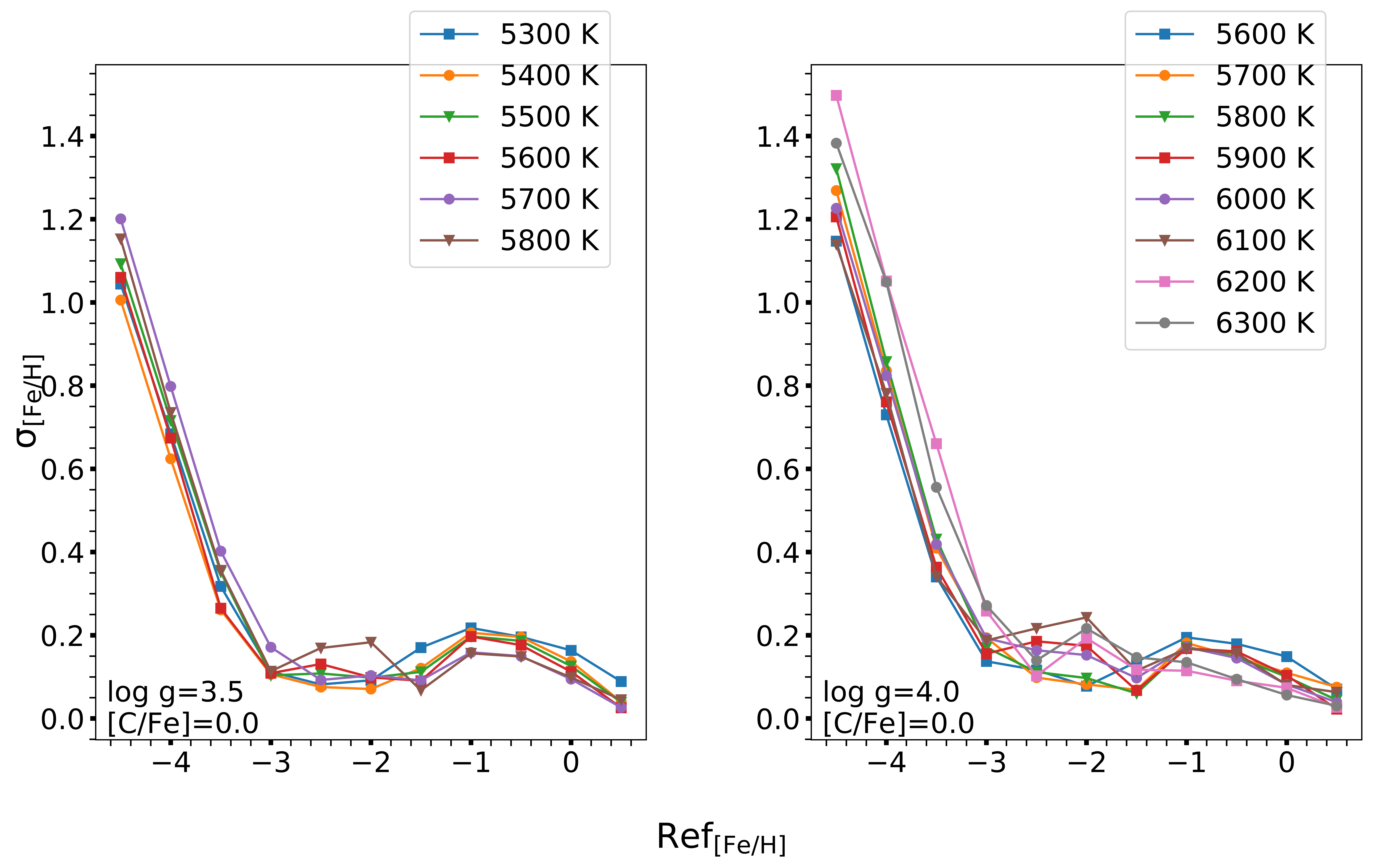}
    \includegraphics[width=0.5\textwidth]{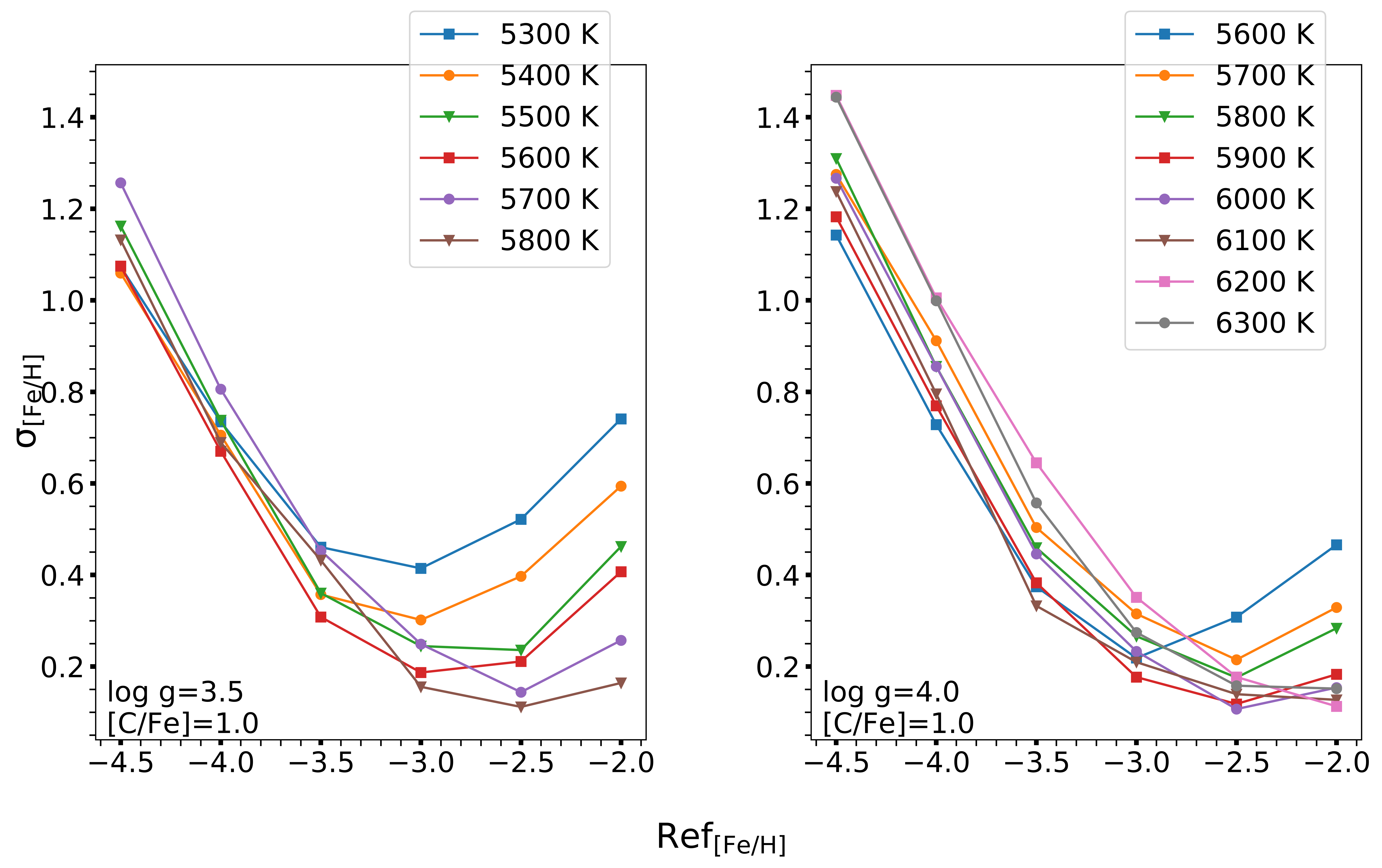}
    \caption{Temperature dependence of the performance of our method is twofold. When $\mathrm{[C/Fe]}\leq0$, the uncertainty is slightly better for lower temperatures up to $\mathrm{[Fe/H]}\sim-3.5$, and above that, it becomes almost independent of $\mathrm{T_{eff}}$. When $\mathrm{[C/Fe]}>0$, the pattern is the same as above for $\mathrm{[Fe/H]}$ up to $\sim-3.5$; whereas, above -3 dex, lower temperatures render higher uncertainties.}
   \label{fig:mfunction_sigma_teff_15mag}
\end{figure}

\begin{figure}
   \centering
   \includegraphics[width=0.5\textwidth]{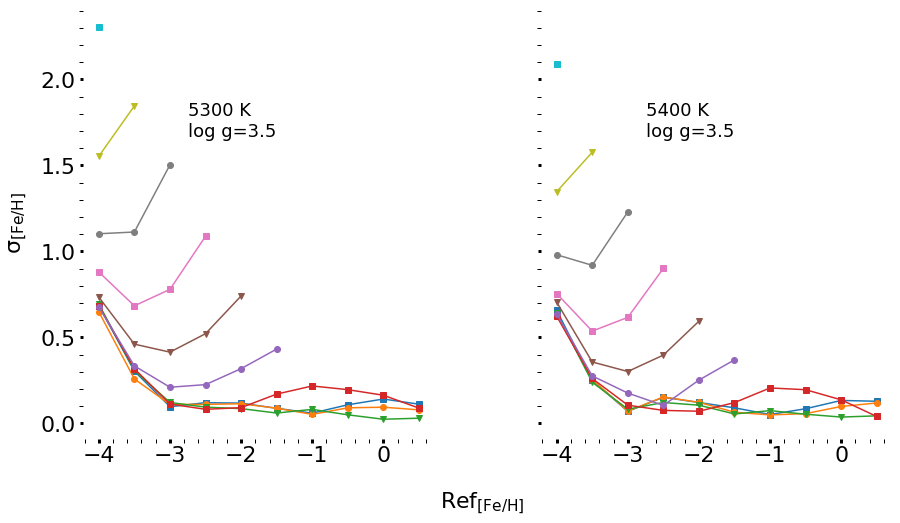}
   \includegraphics[width=0.5\textwidth]{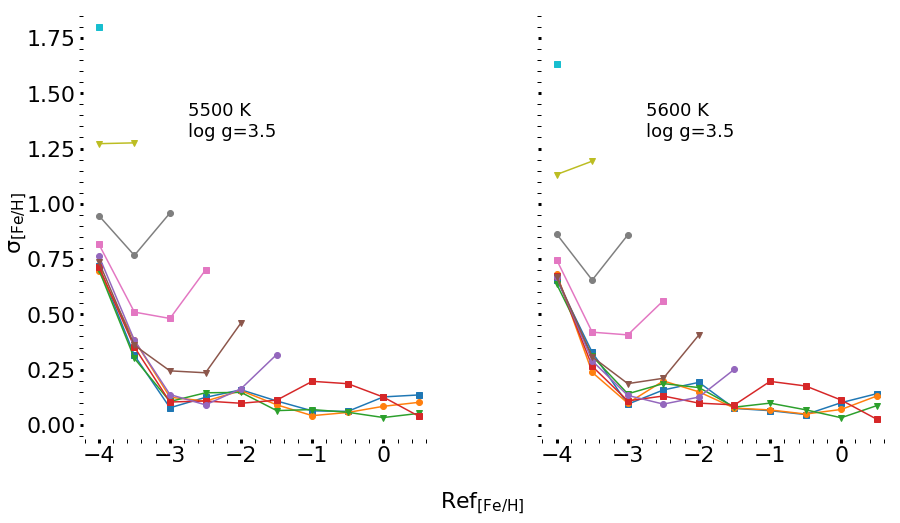}
   \includegraphics[width=0.5\textwidth]{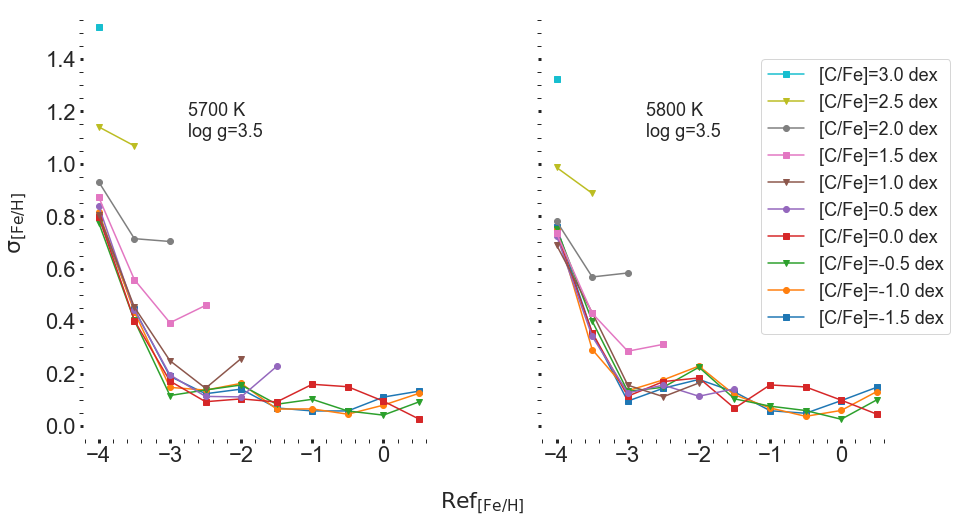}
    \caption{Computation of $\sigma_\mathrm{{[Fe/H]}}$ for all temperatures corresponding to $\mathrm{\log g}=3.5$ dex, and all bins of $\mathrm{[C/Fe]}$. The manner in which the performance of our method depends on the relative carbon abundances relates to whether or not $\mathrm{[C/Fe]\lessgtr0}$ dex: at and below a Solar value, $\sigma_\mathrm{{[Fe/H]}}$ is practically independent of carbon, but above it the uncertainty rises as carbon-enhancement increases. The last effect lessens as temperature rises. The same exercise was performed for all surface gravities of our parameter space, and the results were similar.}
   \label{fig:mfunction_c_impact_15mag}
\end{figure}

The figures in the appendix are described in Sections \ref{sec_mtfr} and \ref{sec_appl}. Figures \ref{fig:classification} and \ref{fig:mfunction_fitting} describe the way we developed our method, whereas Figures \ref{fig:mfunction_sigma_teff_15mag} and \ref{fig:mfunction_c_impact_15mag} detail the dependence of our method on effective temperature and relative carbon abundance. In Table \ref{table:model} we provide the entire dataset that we used to produce Figure \ref{fig:met_relation}, so that the interested readers can apply our method themselves.

\twocolumn

\begin{table*}
\caption{Noise-free $fr\mathrm{_{CaHK/H\beta}}$ and $fr\mathrm{_{G/CaNIR}}$ flux ratios and their respective astrophysical parameters.}
\label{table:model}
\centering
\begin{tabular}{c c c c c c c}
\hline \hline
a/a &     $fr\mathrm{_{G/CaNIR}}$ &    $fr\mathrm{_{CaHK/H\beta}}$ &    $\mathrm{T_{eff}}$ &  $\mathrm{\log g}$ &  $\mathrm{[Fe/H]}$ &  $\mathrm{[C/Fe]}$ \\
& & & (K) & (dex) & (dex) & (dex)\\  \hline\\

0    &  0.353852 &  0.056188 &  4800.0 &   5.0 &    -0.5 &    -0.5 \\
1    &  0.361829 &  0.056728 &  4800.0 &   5.0 &    -0.5 &    -1.0 \\
2    &  0.365205 &  0.056951 &  4800.0 &   5.0 &    -0.5 &    -1.5 \\
3    &  0.337023 &  0.054872 &  4800.0 &   5.0 &    -0.5 &     0.0 \\
4    &  0.288912 &  0.050126 &  4800.0 &   5.0 &    -0.5 &     0.5 \\
5    &  0.187117 &  0.038496 &  4800.0 &   5.0 &    -0.5 &     1.0 \\
6    &  0.122764 &  0.032674 &  4800.0 &   5.0 &    -0.5 &     1.5 \\
7    &  0.086716 &  0.033368 &  4800.0 &   5.0 &    -0.5 &     2.0 \\
8    &  0.324806 &  0.050360 &  4800.0 &   5.0 &     0.0 &    -0.5 \\
9    &  0.333351 &  0.050936 &  4800.0 &   5.0 &     0.0 &    -1.0 \\
10   &  0.336977 &  0.051179 &  4800.0 &   5.0 &     0.0 &    -1.5 \\
11   &  0.305943 &  0.048967 &  4800.0 &   5.0 &     0.0 &     0.0 \\
12   &  0.227209 &  0.042632 &  4800.0 &   5.0 &     0.0 &     0.5 \\
13   &  0.144015 &  0.034670 &  4800.0 &   5.0 &     0.0 &     1.0 \\
14   &  0.097575 &  0.032297 &  4800.0 &   5.0 &     0.0 &     1.5 \\
15   &  0.073156 &  0.034265 &  4800.0 &   5.0 &     0.0 &     2.0 \\
... & ... & ... & ... & ... & ... & ... \\
8963 &  0.471143 &  0.097089 &  5000.0 &   1.5 &    -7.0 &     0.0 \\
8964 &  0.471170 &  0.097088 &  5000.0 &   1.5 &    -7.0 &     0.5 \\
8965 &  0.471178 &  0.097102 &  5000.0 &   1.5 &    -7.0 &     1.0 \\
8966 &  0.471137 &  0.097087 &  5000.0 &   1.5 &    -7.0 &     1.5 \\
8967 &  0.471192 &  0.097109 &  5000.0 &   1.5 &    -7.0 &     2.0 \\
8968 &  0.471005 &  0.097071 &  5000.0 &   1.5 &    -7.0 &     2.5 \\
8969 &  0.470596 &  0.097043 &  5000.0 &   1.5 &    -7.0 &     3.0 \\
8970 &  0.469710 &  0.096964 &  5000.0 &   1.5 &    -7.0 &     3.5 \\
8971 &  0.466643 &  0.096713 &  5000.0 &   1.5 &    -7.0 &     4.0 \\
8972 &  0.458394 &  0.095966 &  5000.0 &   1.5 &    -7.0 &     4.5 \\
8973 &  0.439977 &  0.094247 &  5000.0 &   1.5 &    -7.0 &     5.0 \\
8974 &  0.409192 &  0.090923 &  5000.0 &   1.5 &    -7.0 &     5.5 \\
8975 &  0.368257 &  0.084704 &  5000.0 &   1.5 &    -7.0 &     6.0 \\ \hline
\end{tabular}
\tablefoot{Figure \ref{fig:met_relation}, as well as our model, was produced from this dataset.}
\end{table*}

\end{appendix}
\end{document}